\documentclass[twocolumn]{aastex631}

\usepackage{amsmath}
\usepackage{soul}
\soulregister{\cite}7
\soulregister{\citep}7
\soulregister{\citet}7

\begin{document}

\title{Constraining the Presence of Companion Planets in Hot Jupiter Planetary System Using TTV Observation from TESS}

\author{Zixin Zhang}
\affiliation{School of Physics and Astronomy, Sun Yat-sen University, Zhuhai 519082, China}
\affiliation{CSST Science Center for the Guangdong-HongKong-Macau Great Bay Area, Sun Yat-sen University, Zhuhai 519082, China}

\author{Wenqin Wang}
\affiliation{School of Physics and Astronomy, Sun Yat-sen University, Zhuhai 519082, China}
\affiliation{CSST Science Center for the Guangdong-HongKong-Macau Great Bay Area, Sun Yat-sen University, Zhuhai 519082, China}

\author{Xinyue Ma}
\affiliation{School of Physics and Astronomy, Sun Yat-sen University, Zhuhai 519082, China}
\affiliation{CSST Science Center for the Guangdong-HongKong-Macau Great Bay Area, Sun Yat-sen University, Zhuhai 519082, China}

\author{Zhangliang Chen}
\affiliation{School of Physics and Astronomy, Sun Yat-sen University, Zhuhai 519082, China}
\affiliation{CSST Science Center for the Guangdong-HongKong-Macau Great Bay Area, Sun Yat-sen University, Zhuhai 519082, China}

\author{Yonghao Wang}
\affiliation{Hainan University, China}

\author{Cong Yu}
\affiliation{School of Physics and Astronomy, Sun Yat-sen University, Zhuhai 519082, China}
\affiliation{CSST Science Center for the Guangdong-HongKong-Macau Great Bay Area, Sun Yat-sen University, Zhuhai 519082, China}

\author{Shangfei Liu}
\affiliation{School of Physics and Astronomy, Sun Yat-sen University, Zhuhai 519082, China}
\affiliation{CSST Science Center for the Guangdong-HongKong-Macau Great Bay Area, Sun Yat-sen University, Zhuhai 519082, China}

\author{Yang Gao}
\affiliation{School of Physics and Astronomy, Sun Yat-sen University, Zhuhai 519082, China}
\affiliation{CSST Science Center for the Guangdong-HongKong-Macau Great Bay Area, Sun Yat-sen University, Zhuhai 519082, China}

\author{Baitian Tang}
\affiliation{School of Physics and Astronomy, Sun Yat-sen University, Zhuhai 519082, China}
\affiliation{CSST Science Center for the Guangdong-HongKong-Macau Great Bay Area, Sun Yat-sen University, Zhuhai 519082, China}

\author{Bo Ma}
\affiliation{School of Physics and Astronomy, Sun Yat-sen University, Zhuhai 519082, China}
\affiliation{CSST Science Center for the Guangdong-HongKong-Macau Great Bay Area, Sun Yat-sen University, Zhuhai 519082, China}
\correspondingauthor{Bo Ma}
\email{mabo8@mail.sysu.edu.cn}

\begin{abstract}
The presence of another planetary companion in a transiting exoplanet system can impact its transit light curve, leading to sinusoidal transit timing variations (TTV). By utilizing both $\chi^2$ and RMS analysis, we have combined the TESS observation data with an N-body simulation to investigate the existence of an additional planet in the system and put a limit on its mass. We have developed CMAT, an efficient and user-friendly tool for fitting transit light curves and calculating TTV with a theoretical period, based on which we can give a limit on its hidden companion's mass.
We use 260 hot Jupiter systems from the complete TESS data set to demonstrate the use of CMAT. 
Our findings indicate that, for most systems, the upper mass limit of a companion planet can be restricted to several Jupiter masses. This constraint becomes stronger near resonance orbits, such as the 1:2, 2:1, 3:1, and 4:1 mean motion resonance, where the limit is reduced to several Earth masses.
These findings align with previous studies suggesting that a lack of companion planets with resonance in hot Jupiter systems could potentially support the high eccentricity migration theory. 
Additionally, we observed that the choice between $\chi^2$ or {root mean square (RMS)} method does not significantly affect the upper limit on companion mass; however, $\chi^2$ analysis may result in weaker restrictions but is statistically more robust compared to RMS analysis in most cases. 
\end{abstract}




\section{Introduction}

Hot Jupiters are exoplanets akin to Jupiter in mass and size but distinct due to their short orbital periods of less than 10 days, orbiting very close to their host stars \citep{dawson_origins_2018}. 
The large sizes and short orbital periods make them the most easily studied targets using the transit technique.  
Despite being the first type of exoplanet discovered orbiting a sun-like star \citep{mayor1995-jupitermass}, the origin of hot Jupiters remains a mystery.
Several competing theories attempt to explain their formation: in-situ formation \citep{batygin2016-situ, dawson_origins_2018}, disk migration \citep{goldreich1980-disksatellite, lin1986-tidal,lin1996-orbital, ward1997-protoplanet, kley2012-planetdisk} and high-eccentricity migration \citep{fa1996-dynamical,trilling1998-orbital, wu2003-planet}.

In the high-eccentricity migration scenario, {a} cold Jupiter is launched to eccentric orbits through dynamical interactions with other planets/stars, followed by tidal friction that can shrink and circularize its orbit over a long period. Since this process is usually dynamically hot, hot Jupiters formed through this channel are likely isolated without any resonant companion planets. 
Therefore, the presence of a nearby planetary companion has long been considered evidence against the high eccentricity migration theory. 
Recent research, however, has observed some hot Jupiters with nearby companions, challenging the high eccentricity migration theory \citep{zhu2021-exoplaneta,wu2023-evidencea}.

In multi-planet systems, mutual gravitational perturbations can alter the times of transit. 
Thus, transit timing variations (TTVs) are a powerful tool for characterizing planets in multi-planet systems \citep{holman2005-usea, agol2005-detecting, lithwick_extracting_2012}. For example, it has been used for measuring masses and eccentricities of planets \citep{hadden2014-densities,hadden2016-numerical,hadden2017-kepler}. This is because the TTV amplitude is directly proportional to the perturber's mass \citep{lithwick_extracting_2012}, and the TTV period and phase are related to the perturber's orbital eccentricity. 
Because of the well-known degeneracy between mass and eccentricity, TTVs alone cannot accurately determine the companion planet's mass. Nevertheless, the phases and amplitudes of TTVs can be used to establish upper limits on the masses of companion planetary candidates \citep{xie_transit_2013, wang_transiting_2021}. 

Launched in 2018, the Transiting Exoplanet Survey Satellite \citep[TESS,][]{ricker_transiting_2014} succeeded the Kepler space telescope, aiming to enhance our knowledge of exoplanets through the transit method. 
TESS's wide field of view and high photometric accuracy make it particularly effective for studying hot Jupiters. It can survey nearly the entire sky and detect subtle brightness changes as low as 20 parts per million (ppm) for a magnitude 12 star with a 2-minute exposure. 
These capabilities make TESS an excellent tool for detecting hot Jupiters, significantly advancing exoplanetary research.
Studying TTVs of a large sample of hot Jupiters using TESS data can put additional constraints on the hot Jupiter's nearby companion planet and its formation channel. The short orbital periods and large radii of hot Jupiters make them ideal targets for TTV analysis. With accurate TTV measurements from TESS, we can constrain the presence of nearby companions and potentially shed light on the formation of hot Jupiters.

Here in this paper, we employ TTV analysis and numerical three-body simulations to constrain the presence of additional planets around 260 hot Jupiters using TESS observation data. 
This approach allows us to establish the upper mass limit of potential companion planets and provides insight into whether the hot Jupiter is situated in an isolated system. 
Our research builds upon the foundational work of \citet{wang_transiting_2018,wang_transiting_2021}, who investigated Transit Timing Variations (TTVs) in 39 hot Jupiter systems. Utilizing 127 new transit light curves from two ground-based telescopes, they were able to impose significant constraints on the presence of nearby companions. 
Despite these contributions, the sample sizes in these studies remain relatively modest, and the constraints they impose on the masses of hidden companions are less stringent compared to this study. 
In this work, we also introduce CMAT\citep{zhang_2024_13739647}\footnote{Available at \url{https://github.com/troyzx/CMAT}}, a Python package developed to analyze TTVs and establish the upper mass limit of planetary companions orbiting hot Jupiters.

This paper is structured as follows. Section~\ref{sec:data} provides an overview of the data used in this study. In Section \Ref{sec:method}, we outline our methodology for analyzing TTVs of hot Jupiters and determining the upper mass limits of their companion planets. 
Our results and findings are described in Section~\Ref{sec:result}. 
Comparisons with previous studies and the broader implications for hot Jupiter formation are discussed in Section~ \Ref{sec:discussion}. 
A summary of our findings is provided in Section~\Ref{sec:conclusion}.

\section{Data and Observation}
\label{sec:data}

The data for this study are taken from the TESS observations.
{We select transiting hot Jupiters observed by TESS based on the following criteria: an orbital period of less than 10 days and a mass greater than 0.3~$\mathrm{M_J}$. 
This results in a sample of 260 hot Jupiter systems, the largest sample size used in such a study to date.}

\begin{figure}
	\centering
	\includegraphics[width=\columnwidth]{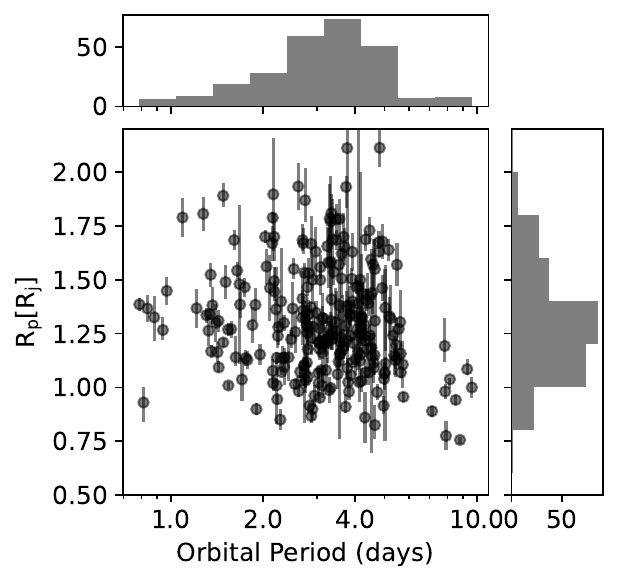}
	\caption{Orbital periods and radii of 260 hot Jupiter systems, the majority of which have orbital periods between 2-6 days and radii between 0.8-1.6 Jupiter radii}
	\label{fig:period_radius}
\end{figure}

\begin{figure}
	\centering
	\includegraphics[width=\columnwidth]{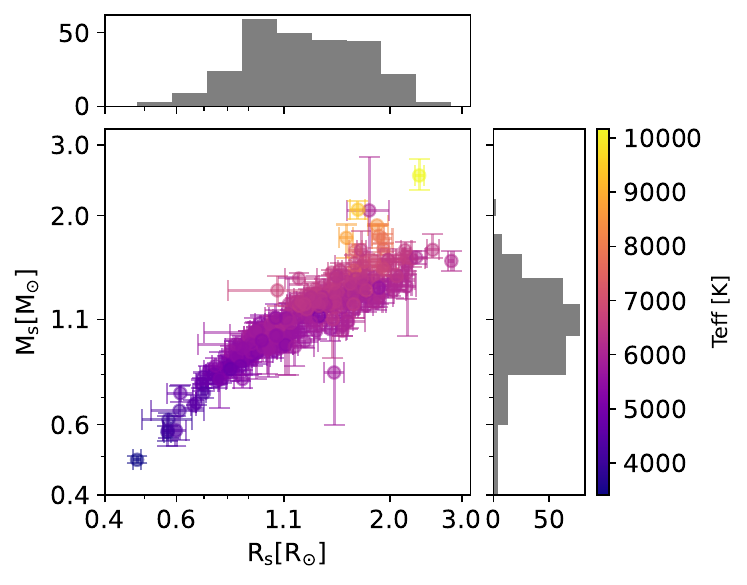}
	\caption{Properties of the host stars in our sample of 260 hot Jupiter systems. }
	\label{fig:star_properties}
\end{figure}

The HJ system parameters, as shown in Figure \ref{fig:period_radius}, reveal that most hot Jupiters have a radius between 0.8 and 1.6 Jupiter radii and an orbital period ranging from 1 to 10 days. 
Further discussions on how the parameters of hot Jupiters can affect the mass of their companions can be found in Section \ref{sec:result}. 
The properties of the host stars in our sample are presented in Figure~\Ref{fig:star_properties}. 
Most of the host stars have masses ranging from 0.8 to 2.0 solar masses, radii ranging from 0.6 to 3 solar radii, and effective temperatures in the range of 5000 to 8000~K.

\begin{figure}
	\centering
	\includegraphics[width=\columnwidth]{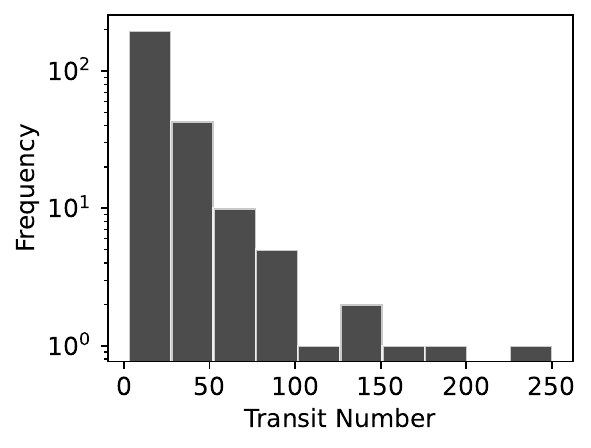}
	\caption{Distribution of transit number for 260 hot Jupiter systems. The majority of systems have transit less than 100, while some of them have more than 200 transits.}
	\label{fig:transit_len}
\end{figure}

{
We use Astroquery \citep{ginsburg2017-astroquery} to retrieve the TESS data products for hot Jupiters generated by the SPOC pipeline \citep{Jenkins16-spoc} from MAST \citep{brasseur2020-mast, https://doi.org/10.17909/t9-nmc8-f686}. 
}
To ensure that we can obtain high-precision measurements of TTV, we included only systems with a minimum of 3 transits covered by TESS observations. 
{Furthermore, we obtained these systems' parameters from the NASA Exoplanet Archive \citep{2013PASP..125..989A}, including stellar mass, planetary mass, and orbital period.} 
Figure~\Ref{fig:transit_len} displays the number of transits in the data we used from TESS observations for our sample of 260 hot Jupiter systems. While the majority of systems have fewer than 100 transits, some exceed this number, with one system having more than 200 transits, namely Qatar-10 b \citep{alsubai2019-qatar}.

{
The presearch data conditioning simple aperture photometry \citep[PDCSAP,][]{2012PASP..124..985S, 2012PASP..124.1000S, 2014PASP..126..100S} light curves from SPOC \citep{2016SPIE.9913E..3EJ} with 2 minute time sampling (``cadence'') are used in this analysis, where common instrumental systematics have been removed. 
To ensure the quality of the light curves, we also remove any outlier data points with a quality flag $>0$. 
Subsequently, we normalized the light curves to account for the varying brightness levels of different stars. 
Following this, we used our CMAT tool to fit the transit light curves and calculate the mid-transit times of HJs, which will be detailed in the next section.
}



\section{Method}
\label{sec:method}
In this section, we present our TTV analysis method. Initially, we evaluate the TTVs by determining the amplitude caused by the companion. Subsequently, precise transit times are calculated from the TESS time series data, and adjustments for TTV are made based on a linear fit of the transit times. A dynamic simulation of the system is conducted to replicate observations and generate simulated TTVs using the transit light curve of the hot Jupiter. Finally, we combine the observed and simulated TTVs ($\mathrm{TTV}_{\mathrm{observed}}$ and $\mathrm{TTV}_{\mathrm{simulated}}$) using $\chi^2$ and RMS analysis to calculate an upper mass limit for the companion. {Figure \Ref{fig:flowchart} illustrates the flowchart of our TTV analysis method.}
{
\begin{figure}
	\centering
	\includegraphics[width=\columnwidth]{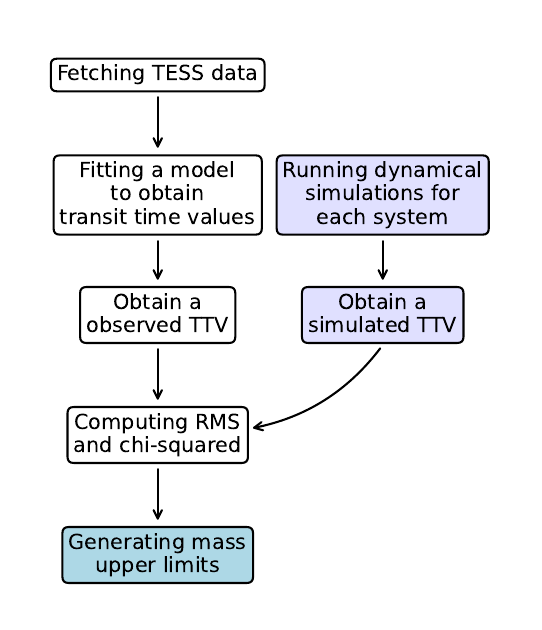}
	\caption{Flowchart of our TTV analysis method.}
	\label{fig:flowchart}
\end{figure}
}

\subsection{TTV from Light-curve Analysis \label{sec:TTV_lightcurve}}
\begin{figure*}
	\centering
	\includegraphics[width=17.8cm]{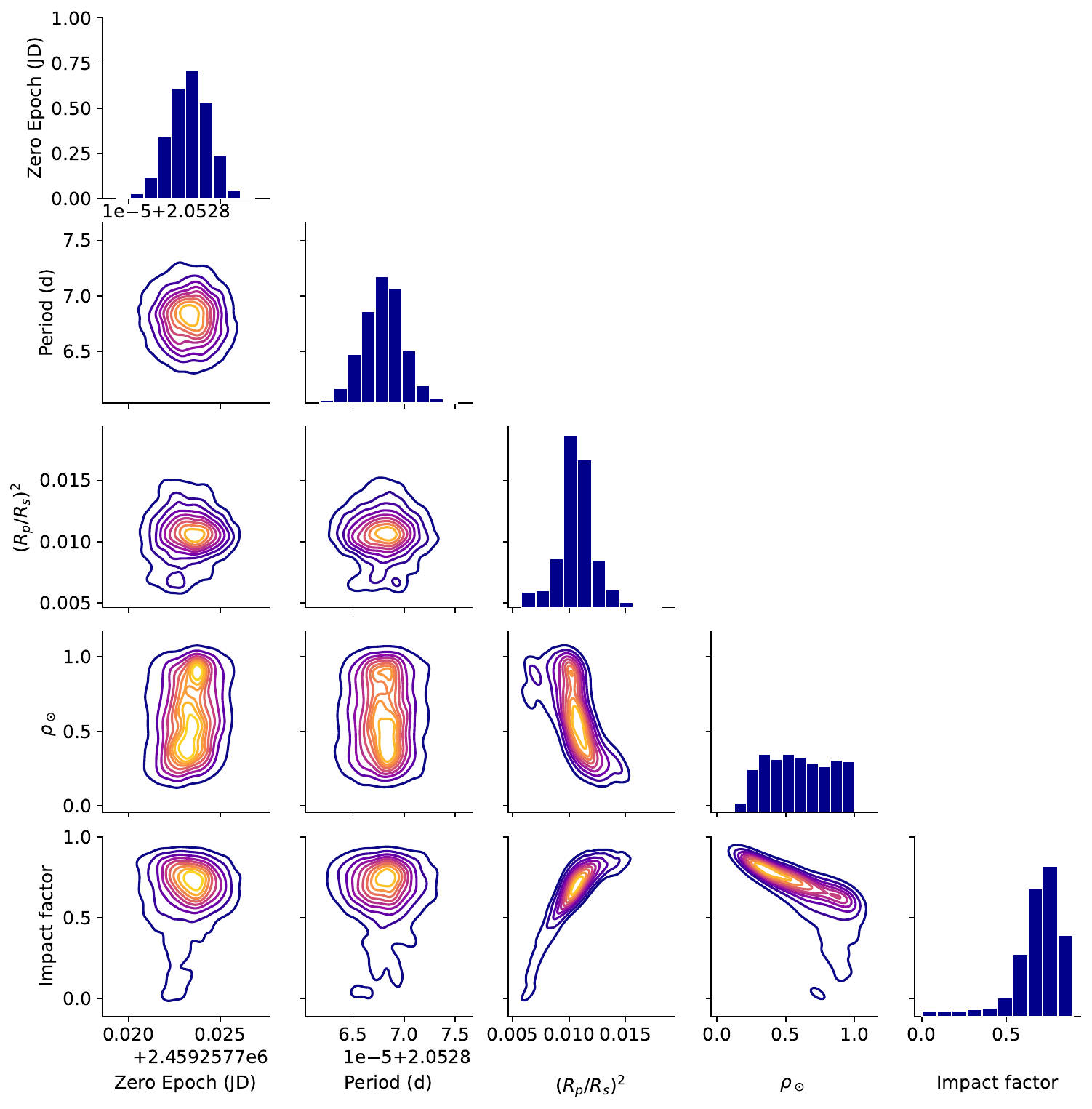}
	\caption{Corner plot derived from the MCMC fitting of WASP-142 b's orbital parameters. This plot specifically illustrates the transit mid-time, orbital period, square of the radius ratio, star density, and impact factor from our result.}
	\label{fig:corner}
\end{figure*}

We derive the transit mid-time data of hot Jupiters using TESS data and Pytransit, similar to our previous studies in \citet{wwq24}. Pytransit is a Python package designed to fit transit light curves \citep{parviainen_pytransit_2015}. It employs the Mandel-Agol model to fit transit light curves and then uses differential evolution and the Markov Chain Monte Carlo (MCMC) method to optimize the fit for 
{stellar and orbital parameters, namely transit midtime, orbital period, planet radius in the unit of star radius, stellar density, and impact factor.}
Figure \Ref{fig:corner} presents the corner plot derived from the MCMC fitting of WASP-142 b's orbital parameters. 

We adapted the Pytransit code to calculate precise transit times and periods for each transit. These values were then aggregated and fitted using a linear model to determine predicted transit times. TTVs were calculated by subtracting the actual transit times from the predicted transit times based on this linear model.

\begin{figure}
	\centering
	\includegraphics[width=\columnwidth]{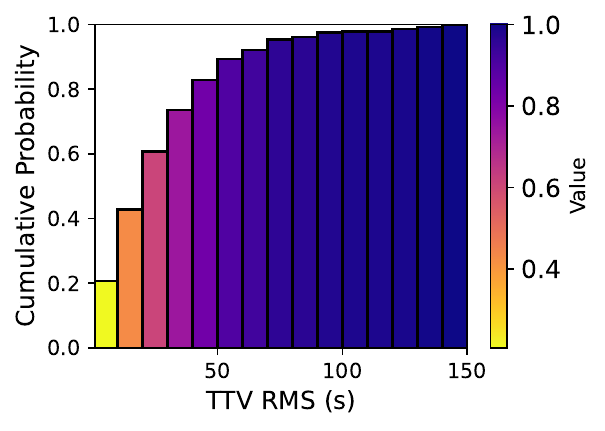}
	\caption{Cumulative probability of Root Mean Square (RMS) of Transit Timing Variations (TTV) calculated from TESS observation data for our 260 hot Jupiter systems sample.}
	\label{fig:cdf}
\end{figure}

Figure \Ref{fig:cdf} displays the cumulative probability distribution of RMS of TTV calculated from TESS observation data for our 260 hot Jupiter systems sample. The graph shows that all of them have an RMS of less than 150 seconds, with over 80\% of the systems having an RMS of less than 50 seconds. 

\begin{figure*}
	\centering
	\includegraphics[width=17.8cm]{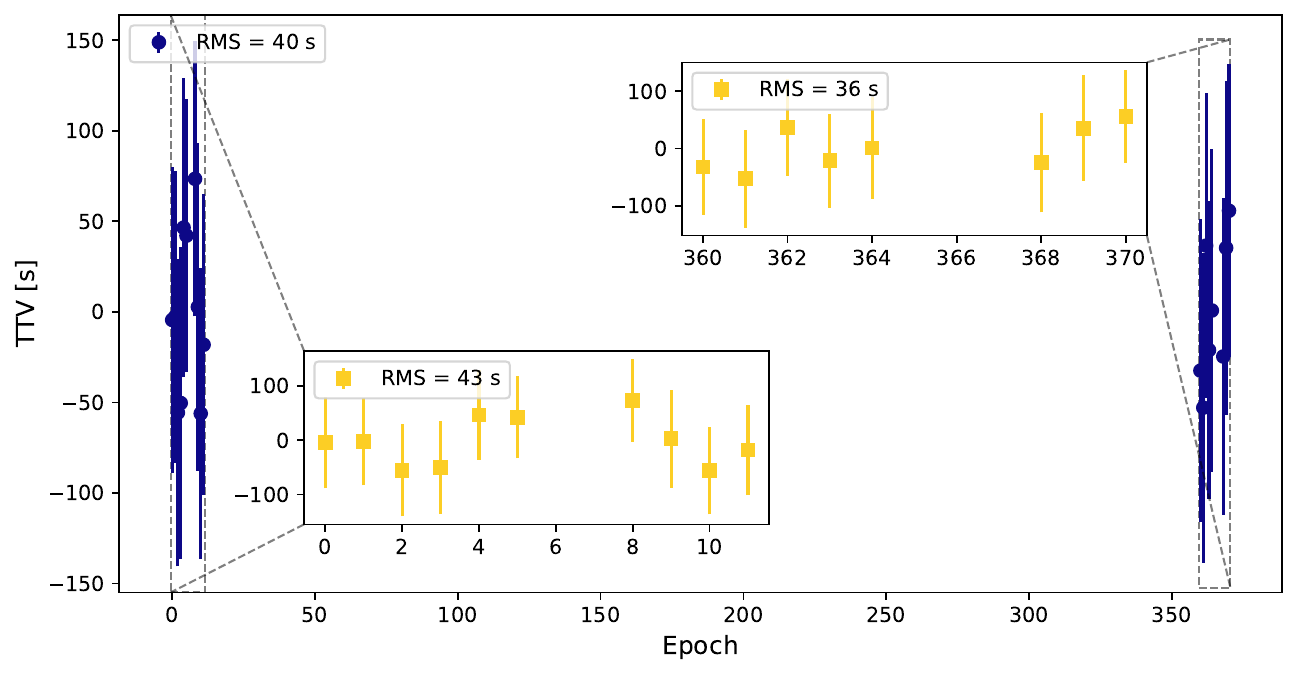}
	\caption{The graph displays the TTV extracted from a linear model of WASP-142 b. The blue line represents the TTV obtained from two separate observation data sets, while the yellow line shows a zoomed-in view for each observation. The total RMS of all observations is 40 seconds, with individual observations having an RMS of 43 and 36 seconds respectively.}
	\label{fig:ttv}
\end{figure*}

WASP-142 b, a hot Jupiter, has an orbital period of 2.05 days and a radius of 1.53 Jupiter radii. It orbits an F-type star, 1.33 times the mass of the Sun \citep{hellier2017-waspsouth}.
Figure \ref{fig:ttv} illustrates the TTV of WASP-142 b calculated by our method. The blue line represents the TTV obtained from two separate observation data sets, while the yellow line shows a zoomed-in view for each observation. The total RMS of all observations is 40 seconds, with individual observations having an RMS of 43 and 36 seconds respectively.
{The graph shows that there are two distinct data sets, and the RMS of each observation is close to one another, as well as the total RMS.}

\subsection{TTV from Dynamical simulation \label{sec:TTV_simulation}}

We employ the rebound package \citep{rein_rebound_2012, rein2015-ias15} to simulate the TTV of a hot Jupiter when an additional companion planet is present in the planetary system. 
The orbit of the additional companion planet is set to be coplanar and circular. 
As discussed by \citet{Agol16}, the coplanar and circular configuration induces smaller TTVs compared with non-coplanar or eccentric orbits. Therefore we can obtain a conservative estimate of the companion planets’ upper mass limits.
We have explored the period ratio of an additional companion planet and hot Jupiter from 0.1 to 4.0. The companion planets' masses are set to vary between 0.01 and $10^5$ Earth masses ($\mathrm{M_{\oplus}}$). 
{The simulation is carried out in a parameter space grid of 300 by 300, where the companion planet mass is in log space.}
We set the system integration time to match the duration of the TESS observation data for each HJ system. 
To encompass all possible coplanar and circular orbital configurations, we have varied the initial orbital phase difference between the HJ and the additional planet at the time of the first HJ transit recorded by TESS from 0 to $2\pi$, using a phase step of $\pi/6$.

\begin{figure}
	\includegraphics[width=\columnwidth]{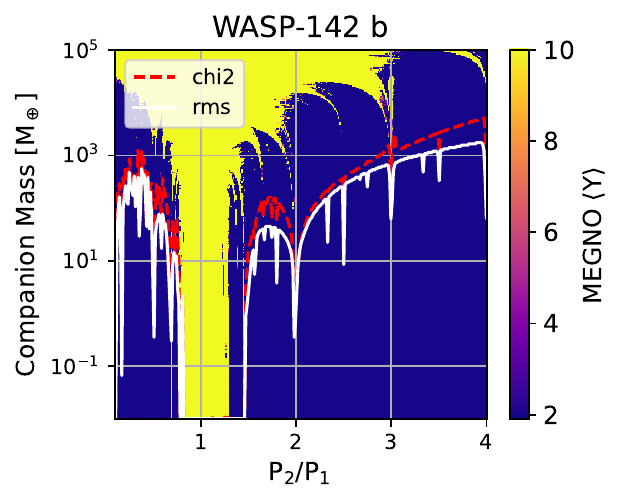}
    \caption{The red dashed line and white line represent the upper mass limit of a possible companion around WASP-142 b, calculated by $\chi^2$ and RMS analysis, respectively. The background color represents {Mean Exponential Growth factor of Nearby Orbits (MEGNO) calculated from the rebound simulation, which helps us assess the dynamical stability of the simulated planetary system. The MEGNO value is a measure of chaos in a dynamical system, with $\langle Y \rangle$ converging to 2 indicating a stable system.}}
    \label{fig:WASP-142b}
\end{figure}

Additionally, we also compute the Mean Exponential Growth factor of Nearby Orbits (MEGNO) \citep{cincotta2003-phase} to assess the dynamical stability of the simulated planetary system, as depicted in Figure~\Ref{fig:WASP-142b}. The MEGNO value is a measure of chaos in a dynamical system, with $\langle Y \rangle$ converging to 2 indicating a stable system. We integrated {each} system for 10000 orbits to test the stability of the system. 

\subsection{Constraints on Additional Planets}
The presence of an additional planet in the transiting hot Jupiter system can perturb the hot Jupiter and generate significant TTV.
To constrain the mass of the additional companion planet, we have employed two different methods, the $\chi^2$ method, and the RMS method, to compare the simulated TTV dataset with the observed TTV dataset. We will proceed to describe the specifics of these two methods next.

The $\chi ^2$ test is a statistical tool that assesses the fit of a model to a dataset. It utilizes the chi-squared distribution, which adds up the squares of independent standard normal random variables. 
The $\chi ^2$ can be expressed as:
\begin{align}
    \chi ^2 = \sum_i\left(\frac{O_i-E_i}{\sigma_i}\right)^2,
\end{align}
where $O_i$ and $\sigma_i$ represent the measured mid-transit time from observation and corresponding uncertainty from Section~\ref{sec:TTV_lightcurve}, and $E_i$ represent the simulated mid-transit time values as described in Section~\ref{sec:TTV_simulation}.

\begin{figure*}
	\centering
	\includegraphics[width=17.8cm]{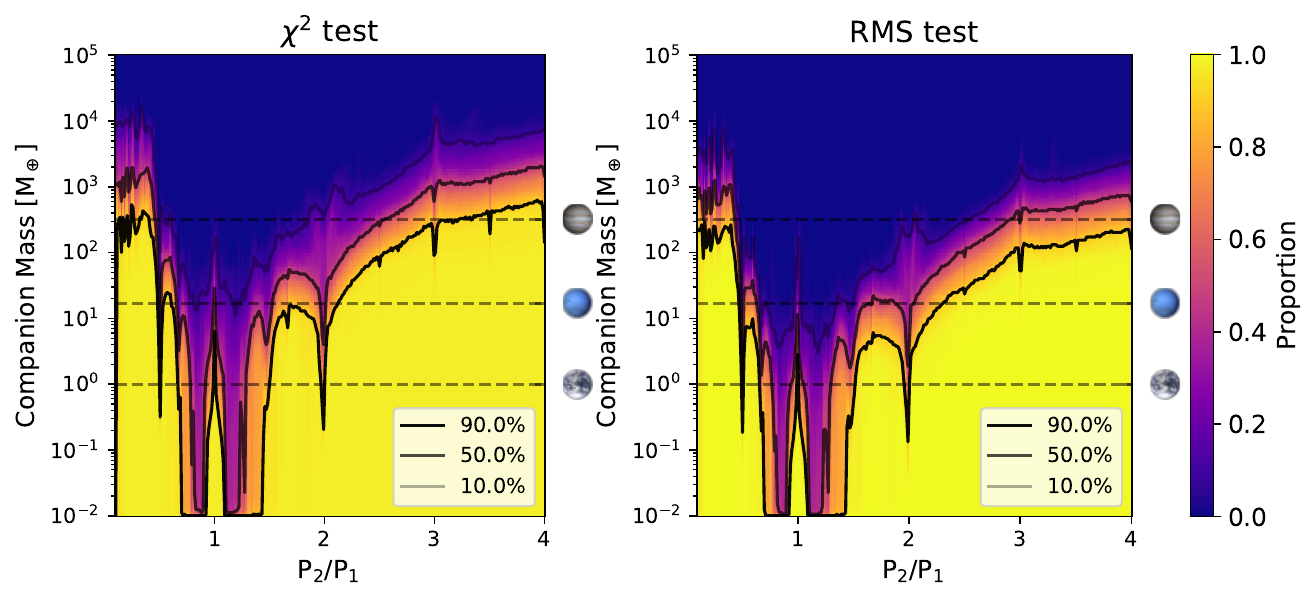}
	\caption{Combined constraints of hidden planet from our TTV analysis of 260 HJ systems. 
 The color of each grid point indicates the percentage of HJ systems within our sample that can hide companion planets with the corresponding planet mass and period ratio. 
 The three solid lines represent equal-percentage lines near $10\%$, $50\%$, and $90\%$, respectively. The left panel shows the results from the $\chi^2$ method, while the right panel displays the results from the RMS test. The masses of Earth, Neptune, and Jupiter are marked with dashed lines. 
}
	\label{fig:combined_result}
\end{figure*}

RMS is a statistical measure of the dispersion of a set of data from its mean. 
The RMS method has been widely used to analyze the TTV observation data of hot Jupiters in the past. For instance, HAT-P-25~b is a hot Jupiter discovered by \citet{quinn2011-hatp25b}. \citet{wang_transiting_2018} have utilized the RMS approach to establish the maximum mass limit of HAT-P-25~b's companion and have ruled out any hidden long-period perturber with $M_p > 0.5, 0.3,$ 
{and}
$0.5 \mathrm{M_\oplus}$ near the 1:2, 2:1, and 3:1 resonances. 
It is calculated by taking the square root of the mean of the squared TTV values. 
The RMS value of TTVs from TESS observation and numerical simulation can be calculated as:
\begin{align}
 	RMS_{\rm obs} = \sqrt{\frac{1}{n}\sum_{i=1}^{n} O_{i}^2}, \\
   	RMS_{\rm sim} = \sqrt{\frac{1}{n}\sum_{i=1}^{n} E_{i}^2}.
\end{align}


After obtaining the TTV data from both observations ($O_{i}$) and simulations ($E_{i}$), we calculated the RMS and $\chi^2$ value for each HJ system. 
The upper mass limit of any companion planet was then established by comparing the simulated TTV data with observed TTV data using either the RMS method or the $\chi^2$ method. 
For each period ratio, we reduced the potential companion's mass until the RMS value from the simulation ($RMS_{\rm sim}$) fell below the RMS value from the TESS observation ($RMS_{\rm obs}$), or the $\chi^2$ value fell below the 3-$\sigma$ critical threshold. 
The result for WASP-142~b is illustrated in Figure~\ref{fig:WASP-142b} as an example. The red dashed line and white solid line indicate, respectively, the upper mass limit of a potential hidden companion around WASP-142~b as determined by $\chi^2$ and RMS analysis. The background color corresponds to MEGNO values calculated from the REBOUND simulations.

{
For seven HJ systems in our sample (HAT-P-59~b, KELT-8~b, KELT-24~b, WASP-62~b, WASP-160~B~b, WASP-170~b, and XO-6~b), none of the simulated TTVs using a three-body model using parameters in our grid can match the observed TTV data. 
There are two possible explanations for this discrepancy. 
One is that we may have underestimated the measurement errors in the observational data, possibly due to unknown systematics. 
The other is that underlying physical processes in the HJ systems may be causing the observed TTVs to deviate from the predictions of a simple three-body model. 
As a result, we are unable to place meaningful constraints on the companion planets using the $\chi^2$ method for these seven systems (see also Figure~\ref{fig:individual_result}), and we encourage further TTV follow-up studies of these seven systems.
}

\subsection{CMAT}

Here we introduce Companion MAss from Transit timing variation (CMAT), a Python package designed to calculate the TTV of hot Jupiter systems and establish the upper mass limit of their hidden planetary companions. We aim to provide a user-friendly tool for determining the upper mass limit of hidden planetary companions by analyzing high-precision lightcurve data. 
CMAT utilizes the MCMC method to accurately compute transit mid-times and TTVs based on input transit time series data. By evaluating $\chi^2$ and RMS values, it can identify the upper mass limit of any hidden planetary companion. The package can also automatically retrieve time series data from exo.MAST. 
To enhance its performance, we integrated multiprocessing into the package via the \texttt{joblib} package \citep{joblib}. Multiprocessing significantly reduces the time required for the time-intensive computation of TTV. For instance, analyzing an observation of 100 transits takes approximately 2 minutes with 96 threads. 
We also plan to incorporate additional features in future updates.

For testing purposes, we used CMAT to analyze 32 sources from \citet{wang_transiting_2021}. Our results align with their findings, demonstrating that CMAT is a dependable tool for analyzing TTV data and establishing the upper mass limit of concealed planetary companions.

\section{Result}
\label{sec:result}
In this study, we have analyzed 260 hot Jupiter systems using our newly developed tool CMAT. Our findings indicate that for most systems, the upper mass limit of a hidden companion can be restricted to several Jupiter masses. This restriction becomes even stronger near resonance orbits such as 1:2, 2:1, 3:1, and 4:1, where we were able to limit the additional companion's mass to just a few Earth masses. 

Though some MMR or near-MMR systems have been discovered \citep{quinn2019-nearresonance}, systematic searches by \citet{lissauer2011-architecture} and \citet{fabrycky2014-architecture} indicate that among the multiple planetary systems discovered, the vast majority are neither in nor near low-order mean-motion resonance systems. This aligns with our study's findings, which show a sharp decrease in the upper mass of a companion planet near resonance, indicating a relatively low occurrence rate at these locations.
These findings align with previous studies suggesting that a lack of companion planets with resonance in hot Jupiter systems could potentially support the high eccentricity migration theory. 
Additionally, we observed that the choice between $\chi^2$ or the RMS method does not significantly affect the upper limit on companion mass. 
However, $\chi^2$ analysis usually provides slightly weaker restrictions for the upper mass limit of companion planet compared to the RMS analysis but is statistically more robust.

We present the combined results for all our targets in Figure \ref{fig:combined_result}, which can be seen as a fold of Figure~\ref{fig:WASP-142b} for all 260 hot Jupiter systems. 
In Figure~\ref{fig:WASP-142b}, each grid point's color indicates the percentage of hot Jupiters in our sample that are capable of hiding companion planets with corresponding mass and period ratio. 
The solid lines on the plot indicate equal-percentage levels at around $10\%$, $50\%$, and $90\%$, respectively. 
The left panel displays the statistical results from the $\chi^2$ method, and the right panel shows the results from the RMS method.
Our results indicate that, for half of hot Jupiter systems in our sample, we can limit the mass of a hidden companion planet to just a few Earth masses near the 1:2 and 2:1 MMR. 
It is important to note that our dynamic simulation was based on a coplanar and circular hypothesis. This particular configuration results in smaller TTVs compared to non-coplanar or eccentric orbits, as discussed by \citet{agol2005-detecting}. Consequently, this approach yields a more conservative estimate of upper mass limits for the hidden companion planets. However, according to \citet{xie2016-exoplanet}, most multi-planet systems are coplanar and circular, aligning with our assumption. 

We have found that specific systems in our sample, such as WASP-47 b, WASP-57 b, and WASP-60 b (see Figure~\ref{fig:individual_result}), do not exhibit a significant reduction in the upper mass limit of a companion planet near resonance. 
This can be mainly attributed to the insufficient number of transits available for analysis for these targets, which results in a weaker constraint on the maximum mass of the potential companion planet.
When more observation data are available in the future, we will be able to impose more stringent constraints on the mass of hidden companion planets in these systems. 

We seek to find correlations between the upper mass limits of hidden companion planets with the HJ parameters in Figure~\ref{fig:12_21_31}. 
The primary planet's orbital period ($P$) and radius ($R_p$) are divided into 15 $\times$ 5 bins, with logarithmic spacing for the orbital period. 
The color coding indicates the {median} maximum mass limit in Earth/Jupiter mass for the hidden companions near resonance ratios of 1:2, 2:1, and 3:1 (from top to bottom). 
{
We have calculated the Pearson correlation coefficients between the companion's upper mass limits near the resonance ratios and the HJ's radii and orbital periods.
The results indicate that no significant correlation was found within the currently available dataset. 
However, there is a lack of HJs in the top-right corner bins of Figure~\ref{fig:12_21_31}, and future data in these bins could potentially alter the correlation results.
}

{
\begin{figure*}
	\centering
	\includegraphics[width=17.8cm]{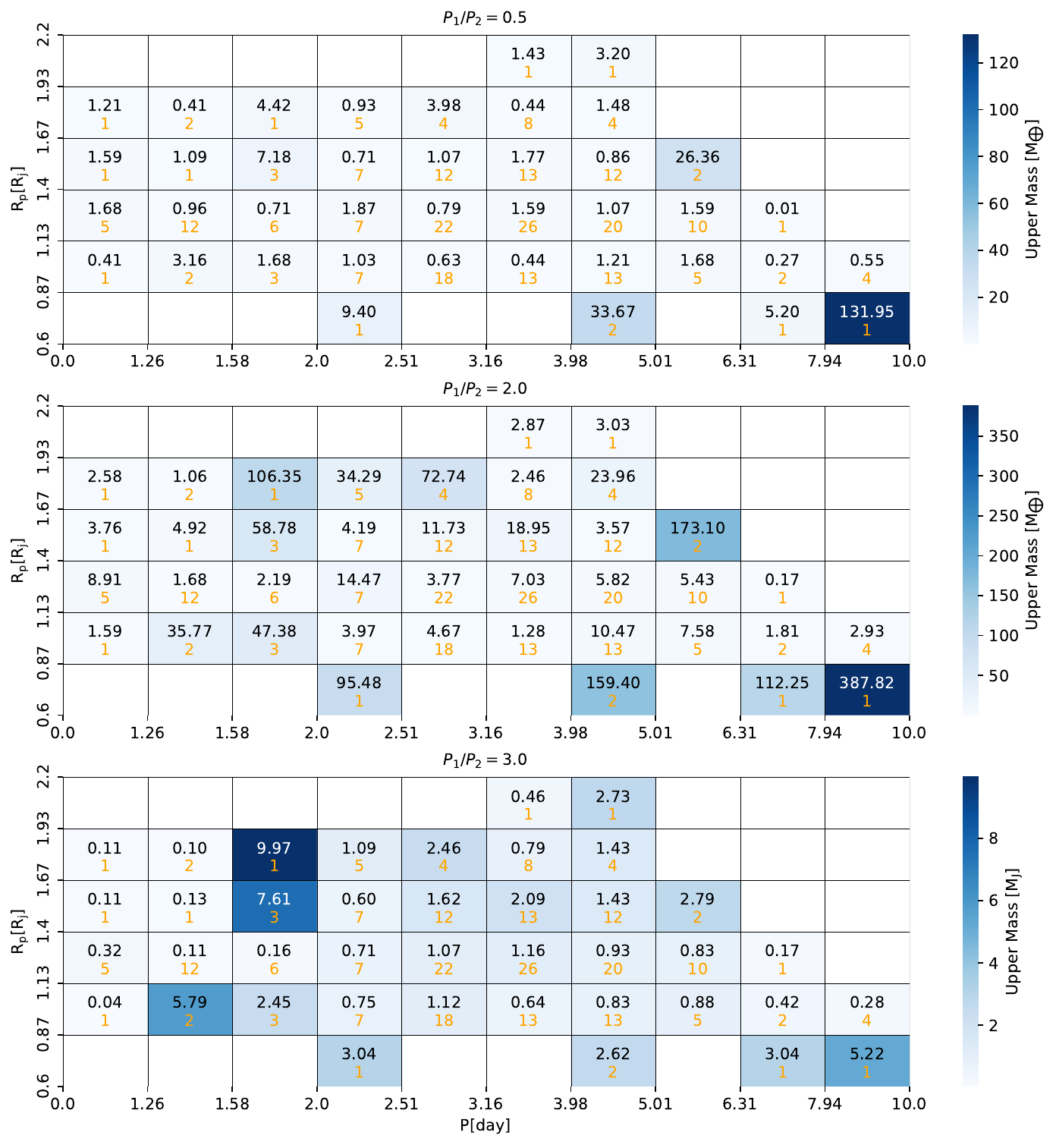}
	\caption{The median upper mass limit of the hidden companion near resonances 1:2, 2:1, and 3:1 is expressed in Earth masses. The orbital period ($P$) and radius ($R_p$) of the primary planet are categorized into 15 $\times$ 5 bins, with the bins for the orbital period logarithmically spaced. {The yellow numbers represent the number of systems in each bin.} The color coding indicates the {median} maximum mass limit of the hidden companion.}
	\label{fig:12_21_31}
\end{figure*}
}

Our results provide valuable insights into the dynamics of hot Jupiter systems and their hidden companions, contributing to the ongoing debate on the origin of hot Jupiters.

\section{Discussion}
\label{sec:discussion}
\subsection{Comparison with Previous Studies}

In this subsection, we compare our findings with those of previous studies utilizing TTV methods to investigate companion planets near hot Jupiter.

\citet{holman2006-transit} launched the Transit Light Curve (TLC) project to create a collection of precise transit photometry data. 
The goal is to improve system parameters and detect variations in transit times and light-curve shapes that could suggest the presence of extra planets \citep{chan2011-transit, fernandez2009-transit, holman2007-transit, holman2007-transita, winn2007-transit, winn2007-transita, winn2007-transitb, winn2007-transitc, winn2008-transit, winn2008-transita, winn2009-transit, winn2009-transita}.
In one of the TLC series papers, \citet{carter2011-transit} used the TTV method {on GJ 1214b} to rule out the presence of a companion planet with a mass exceeding 0.004 and 0.01 $\mathrm{M_\oplus}$ near 1:2 and 2:1 resonances, respectively. 
\citet{gibson2010-transit} studied the TTV of HAT-P-3~b and established upper mass limits of 0.33 and 1.81 Earth masses for a hypothetical planetary companion in 1:2 and 2:1 resonances, respectively. 
By comparison, we have improved upon their constraint by taking upper mass limits down to 0.1 Earth masses at both resonance positions shown in Section~\ref{sec:result}. 
The exploration of Kepler data by \citet{steffen2012-kepler} found no significant TTV signal of a nearby companion planet in 63 Kepler hot Jupiter systems. Nevertheless, a photometric analysis successfully excluded companions with masses ranging from about 2/3 to 5 times that of Earth.

The Transit Monitoring in the South project \citep[TraMoS,][]{hoyer2012-transit} did not detect any TTV RMS variations exceeding 1 minute over a span of 3 years for the hot Jupiter WASP-5 b, and ruled out the existence of planets with masses greater than 5 $\mathrm{M_\oplus}$ and 2 $\mathrm{M_\oplus}$ near the 1:2 and 2:1 resonances, respectively. 
Their subsequent study of WASP-4~b \citep{hoyer2013-tramos} excluded companions with masses exceeding 2.5 and 1.0~$\mathrm{M_\oplus}$ near the 1:2 and 2:1 resonances, respectively. 
Their most recent study of WASP-19~b \citep{cortes-zuleta2020-tramosa} ruled out companions with masses greater than 0.26 and 2.8 $\mathrm{M_\oplus}$ near the 1:2 and 2:1 resonances, respectively. Our study has narrowed down these limits in most cases, reaching as low as 0.39 $\mathrm{M_\oplus}$ and 1.35 $\mathrm{M_\oplus}$ near these resonances, except for the 2:1 resonance of WASP-19~b.

The TEMP project \citep{wang_transiting_2018} aims to study TTV signals of hot Jupiter systems using ground-based telescopes. 
Using data from ground-based telescopes, they have excluded companions with masses above 0.6 and 0.4 $\mathrm{M_\oplus}$ for HAT-P-29~b \citep{wang_transiting_2018}, and 0.5 and 0.3~$\mathrm{M_\oplus}$ for HAT-P-25 b \citep{wang2018-transiting} near the 1:2 and 2:1 resonances, respectively. 
Our study only narrows down this limit to tens of Earth masses near these resonances for both systems due to the small number of transits observed by the TESS mission. 
The TEMP team also investigated the HAT-P-33 b system \citep{wang2017-transiting}, ruling out a perturber with a mass exceeding 0.6 and 0.3~$\mathrm{M_\oplus}$ near the 1:2 and 2:1 resonances, respectively. 
With an adequate number of transits, we have managed to refine this limit to as low as 0.16 and 1.09 $\mathrm{M_\oplus}$ near these resonances. 
In their most recent research, the TEMP team analyzed 39 hot Jupiter systems, excluding companions with masses greater than $0.39-5.0\ \mathrm{M_\oplus}$ and $0.69-6.75\ \mathrm{M_\oplus}$ near the 1:2 and 2:1 resonances \citep{wang_transiting_2021}. 
Our work has further enhanced this constraint by narrowing the lower end down to as small as 0.05~$\mathrm{M_\oplus}$ near these resonances. 

In summary, our study has expanded the sample size compared to previous studies on Transit Timing Variations (TTV) of hot Jupiters and generally establishes lower upper mass limit values compared to prior research. With the largest sample size to date, our study enables us to provide the most robust and statistically significant assessment of the upper mass limit for hidden planetary companions in hot Jupiter systems.

\subsection{Eccentricity and Inclination}

Since multi-planet transiting systems usually have nearly circular orbits and small mutual inclinations, we have used coplanar approximation to put constraints on additional companion planets in the hot Jupiter system in Section~\ref{sec:TTV_simulation}. 
However, it is important to note that TTV signal can be signiﬁcantly ampliﬁed for hierarchical planetary systems with substantial orbital inclinations and/or for an eccentric transiting planet with the anti-aligned orbit of the planetary companion \citep{holman2005-usea, nesvorny2009-transit, Agol16, kubiak2023-ttv}. 
These studies indicate that our reliance on coplanar and circular assumptions may lead to an underestimation of the TTV effects induced by an eccentric and/or inclined hidden companion, consequently resulting in a more conservative upper mass limit estimate. 
If non-coplanar and eccentric orbits are adopted for the concealed companion planet in the hot Jupiter system, we anticipate a decrease in the upper mass limits depicted in Figure~\ref{fig:combined_result}.
In our future work, we plan to consider both inclination and eccentricity to explore the possibility of non-coplanar or non-circular-orbital companions.

\subsection{HJs with Known Companion Planets}
{
Several HJ systems in our sample have known companion planets, including WASP-47, WASP-84, and WASP-132 \citep{Bryant22, Maciejewski23, Grieves24}. We show their results in Figure~11. 
We find all four known companion planets are sitting below the upper limits derived using simple three-body simulations presented in Section~\ref{sec:method}, which in part validates the effectiveness of our method. Thus, we have kept these three HJ systems in our group study presented in Figure~\ref{fig:combined_result} and Figure~\ref{fig:12_21_31}.
However, we do want to note here that our study assumes a simple three-body model. 
And for systems like WASP-47, where there is one star and at least three known planets, our method does not necessarily apply. 
A four- or five-body model is required to put more precise constraints on additional unknown companion planets in these systems, which is beyond the scope of this work. 
}

{
Among these three HJ systems, \citet{Becker15} have detected a TTV variation of WASP-47~b with an amplitude of $0.63\pm0.10$~mins and a period of $\sim53$~days using data from the Kepler mission. \citet{Bryant22} and \citet{Nascimbeni23} have studied the TTVs of WASP-47~b using TESS and CHEOPS data.
Similar to the study by \citet{Bryant22}, we can not confirm this variation using TESS data alone due to the larger timing noise (about 1 minute) and short time coverage (around 20~days). 
}
%

\begin{figure*}
	\centering
	\includegraphics[width=17.8cm]{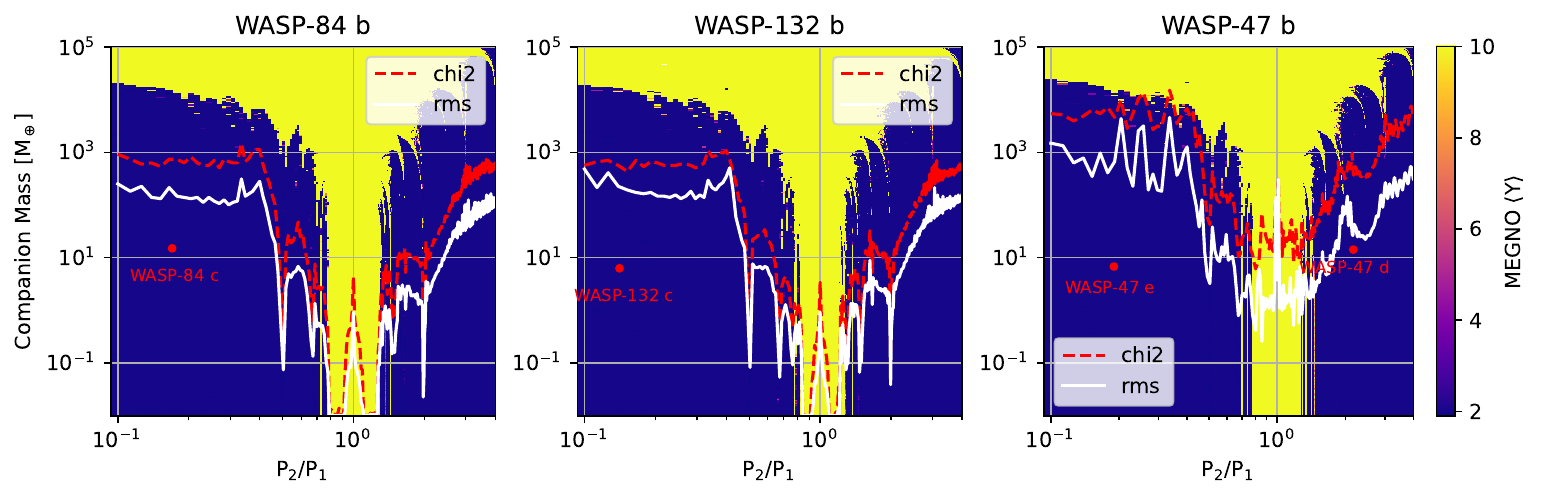}
	\caption{Same as Figure~\ref{fig:ttv}, but for HJ Systems with known companion planets, namely WASP-84 b, WASP-132 b and WASP-47 b.}
	\label{fig:multiple_planet_system}
\end{figure*}

\subsection{Formation Theories of Hot Jupiters}

We will discuss the implications of our results for the study of hot Jupiter formation theories in this section. 
Through the TTV analysis, hidden planetary companions of hot Jupiters can be detected or their existence ruled out. And the constraints become more stringent near resonance orbits. Thus, our findings can be used to constrain theoretical models used to generate hot Jupiter systems. 
Giant planets are more likely to form at larger separations, beyond the ice line of the star, where dust production is enhanced, before migrating inwards \citep{kokubo2002-formation}. 
Studying the existence and distribution of planetary companions in hot Jupiter systems can offer valuable insights into the migration mechanisms of these planets, which are a key aspect of their formation and evolution \citep{wu2023-evidencea, zink23}. 
{Several popular migrating scenarios proposed by previous studies include high eccentricity migration and disk migration.}
In the high eccentricity migration scenario, planet-planet scattering \citep{fa1996-dynamical, chatterjee2008-dynamical,beauge2012-multipleplanet}, Kozai–Lidov effect \citep{wu2003-planet, naoz2012-formation,petrovich2016-warm,vick2019-chaotic}, and secular dynamics \citep{wu2011-secular, petrovich2015-hot} have been invoked to circularize and shrink the orbits of hot Jupiters. 

Resonance chains could be common, as giant planets may have captured terrestrial planets in such chains during migration. Multiple {systems} have been found in this type of resonance chain \citep{tinney2006-resonant}, supporting the disk migration theory. 
If a hot Jupiter has migrated to its current position through high eccentricity migration, there should be no/few companions nearby \citep{mustill2015-destruction}. 
Therefore, detecting a high percentage of hidden companions near the resonance orbits of hot Jupiters would favor the disk migration theory over the high eccentricity migration theory, or vice versa. 
Currently, several population synthesis studies have been carried out to study the formation of exoplanets \citep{mordasini2009-extrasolara, mordasini2009-extrasolarb, alibert2011-extrasolar, mordasini2012-extrasolara,  emsenhuber2021-new, emsenhuber2021-newa, schlecker2021-new, burn2021-newa, schlecker2021-newa, mishra2021-new}. Results from similar population synthesis studies in the future for HJs can be used to directly compare with our statistical results.

Here, we propose two ways to utilize our findings for testing planetary formation models. 
The first one involves generating a synthetic population of hot Jupiters using a planetary population synthesis model. 
We then identify those with hidden companions in the vicinity of MMR. By comparing their mass distribution with our results, we can assess the feasibility of the HJ formation model.
The second one also involves generating a synthetic population of hot Jupiters. 
We can then perform a statistical analysis of the synthetic population to obtain the percentages of hot Jupiters with companion planets above the three percentage lines shown in Figure~\ref{fig:combined_result}. 
If any of the three percentages from the synthetic population is larger than the corresponding percentage (90$\%$, 50$\%$, or 10$\%$) of the three solid 
lines, which means there are too many HJ systems with companion planets from theory than from observation, we can then infer that the formation model is inconsistent with observations.

A potential source of statistical bias arises from the fact that the discovery of hot Jupiters may favor systems with weak interactions, resulting in smaller transit timing variations (TTVs). Large TTVs might hinder the detection of hot Jupiters using detection algorithms such as the Box-fitting Least Squares (BLS) method. However, given the large transit depths of hot Jupiters, this bias is expected to be minimal. Future studies on TTVs of smaller planets should consider and adjust for this potential bias to ensure accuracy.

\section{Summary and Conclusion}
\label{sec:conclusion}

The study of hot Jupiters and their companion planets is an active research field in exoplanet science.
Here we investigate the existence of additional planets in 260 hot Jupiter systems by analyzing transit timing variations observed by the TESS mission. 
The use of our newly developed CMAT tool, which combines observation data with numerical N-body simulations using $\chi^2$ and RMS analysis, has enabled us to provide more stringent constraints on the mass of the hidden companions than previous studies. 
This is particularly evident near resonance orbits, where we have been able to limit the companion planet mass to just a few Earth masses. 
These findings not only enhance our understanding of the dynamics of hot Jupiter systems but also provide valuable insights into the processes of planet formation and migration. The absence of resonant companion planets in hot Jupiter systems, as revealed by our study, lends support to the high eccentricity migration theory, contributing to the ongoing debate on the origin of hot Jupiters.
We also note that $\chi^2$ analysis, while less restrictive, offers a statistically robust approach compared to RMS analysis. 

Our work has pushed the boundaries of what is currently known about hot Jupiter systems and their hidden companions. 
With more high precision transit time data becoming available in the future, such as from PLAnetary Transits and Oscillation of
stars mission \citep[PLATO,][]{Rauer2014}, the China Space Station Telescope \citep[CSST,][]{zhan11}, and the Earth 2.0 mission \citep[ET,][]{ge22}, we expect more detailed measurements leading to better comprehension of planetary system formation and evolution within our galaxy as well as beyond it.

We plan to extend our TTV study from hot Jupiters to short-period transiting brown dwarfs (BDs) to further investigate the differences between the hot Jupiter and brown dwarf populations, as highlighted by researchers such as \citet{Grether06} and \citet{Ma14}. For instance, \citet{Bowler20} argued that high-mass brown dwarfs predominantly form similarly to stellar binaries, based on a comparison of the mass-eccentricity distributions of brown dwarfs and giant planets. Since TTV studies can help reveal the dynamic properties of brown dwarf systems, we believe that expanding TTV analysis to transiting BDs can provide additional observational evidence for distinct formation mechanisms.
Therefore, we encourage our colleagues to continue monitoring the transits of both hot Jupiters and transiting brown dwarfs to enhance our understanding of these celestial objects.

\section*{Acknowledgements}

We express our gratitude to the TESS mission for its outstanding contribution to exoplanet science. The precise transit-time measurements from TESS were instrumental in our study of TTVs in hot Jupiters. We thank NSFC grants 12073092, 12103097, and 12103098, the National Key R$\&$D Program of China (2020YFC2201400), as well as the science research grants from the China Manned Space Project (No. CMS-CSST2021-B09) for their financial support. Additionally, we appreciate the valuable assistance of the NASA Exoplanet Archive, which granted us access to a vast amount of observational data and resources that greatly aided our research efforts. We also thank open-source software developers for their contributions to the scientific community. 

\facilities{TESS}

\software{PyTransit, CMAT, {REBOUND, Astroquery}}


\bibliographystyle{aasjournal}
\bibliography{main.bib} 



\appendix

\section{Result of Individual Target}
\begin{figure*}[htb!]
    \centering
    \includegraphics[page=1, width=0.93\textwidth]{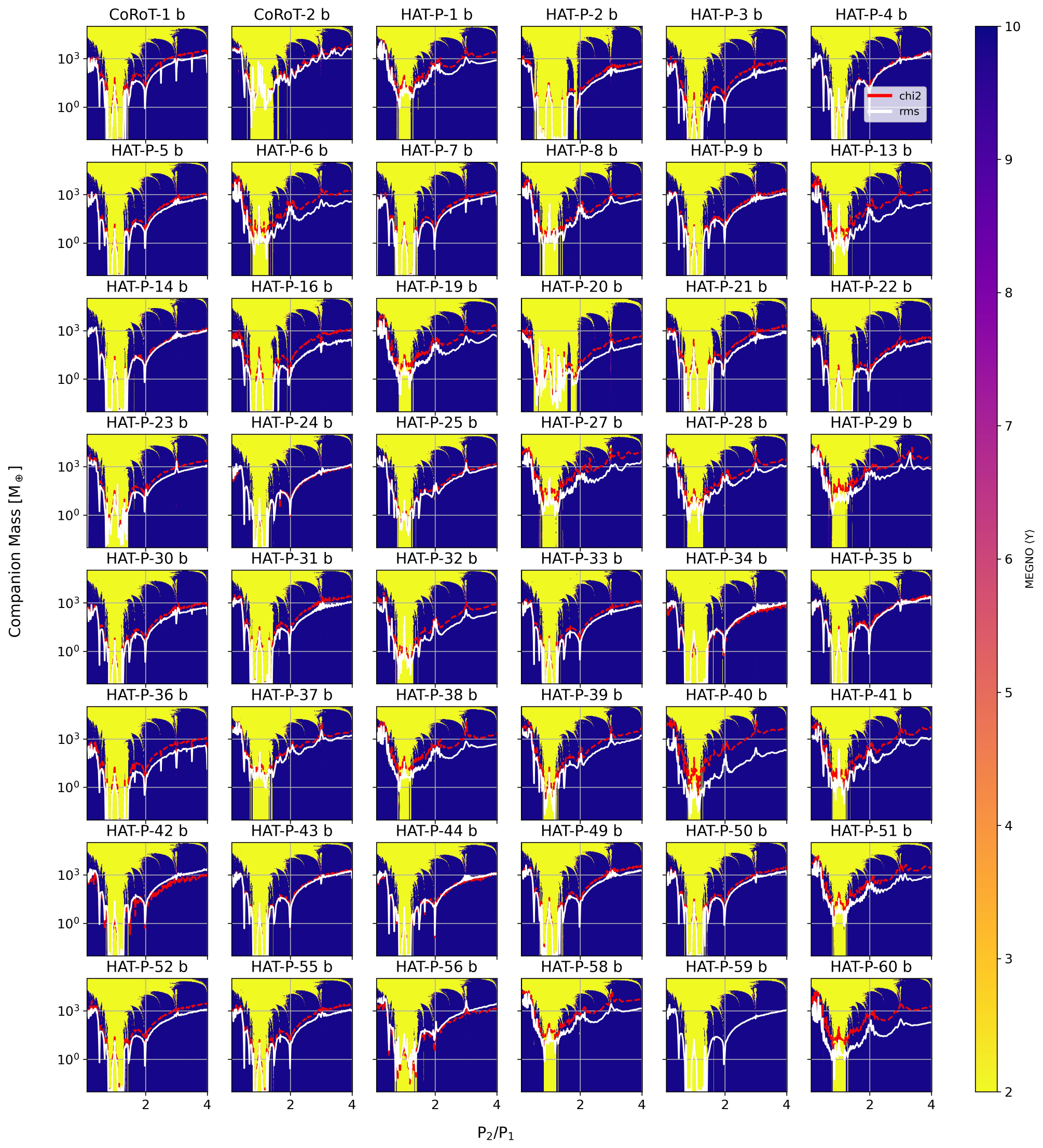}
    \caption{Upper mass limit of companion planets and MEGNO for 260 HJs in our sample.}
    \label{fig:individual_result}
    \figurenum{A1}
\end{figure*}

In this appendix, we present the individual upper mass limit of hidden companions for 260 hot Jupiter systems. The upper mass limit is calculated using $\chi^2$ and RMS analysis. The results are shown in Figure \ref{fig:individual_result}.

\restartappendixnumbering

\begin{figure*}[htb!]
    \centering
    \includegraphics[page=2, width=0.95\textwidth]{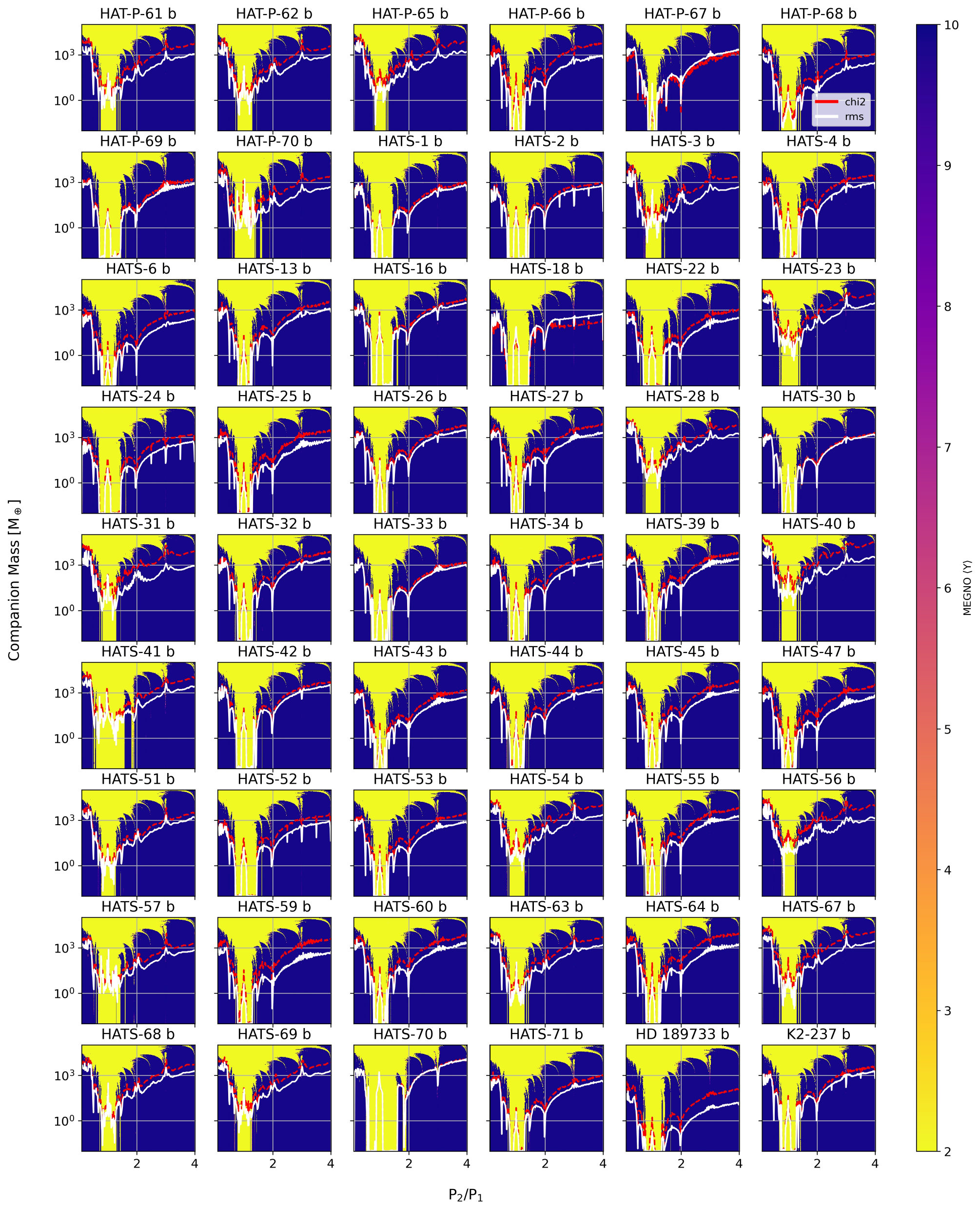}
    \caption{Upper mass limit of companion planets and MEGNO for 260 HJs in our sample. (Continued)}
    \figurenum{A1}
\end{figure*}

\begin{figure*}[htb!]
    \centering
    \includegraphics[page=3, width=0.95\textwidth]{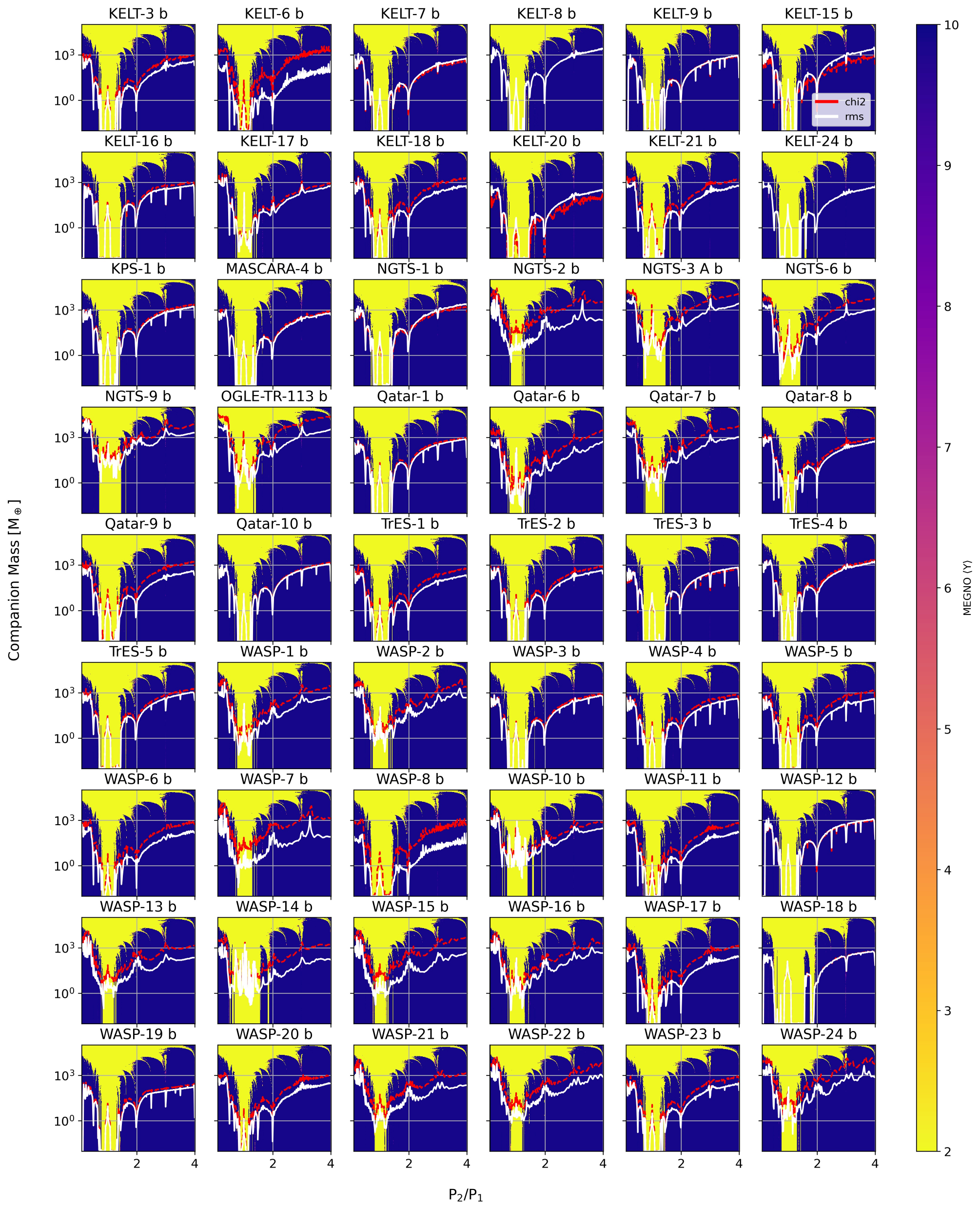}
    \caption{Upper mass limit of companion planets and MEGNO for 260 HJs in our sample. (Continued)}
    \figurenum{A1}
\end{figure*}

\begin{figure*}[htb!]
    \centering
    \includegraphics[page=4, width=0.95\textwidth]{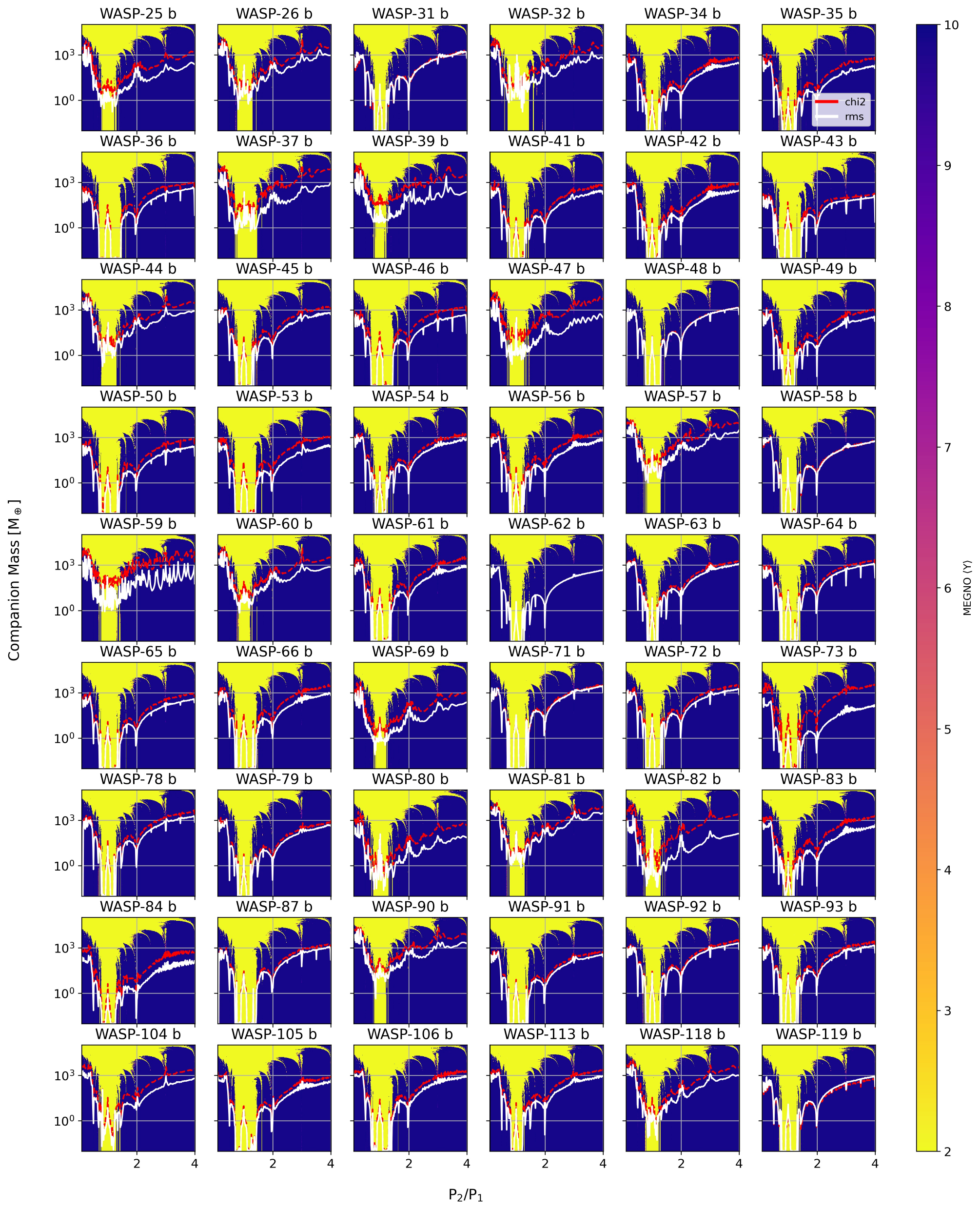}
    \caption{Upper mass limit of companion planets and MEGNO for 260 HJs in our sample. (Continued)}
    \figurenum{A1}
\end{figure*}

\begin{figure*}[htb!]
    \centering
    \includegraphics[page=5, width=0.95\textwidth]{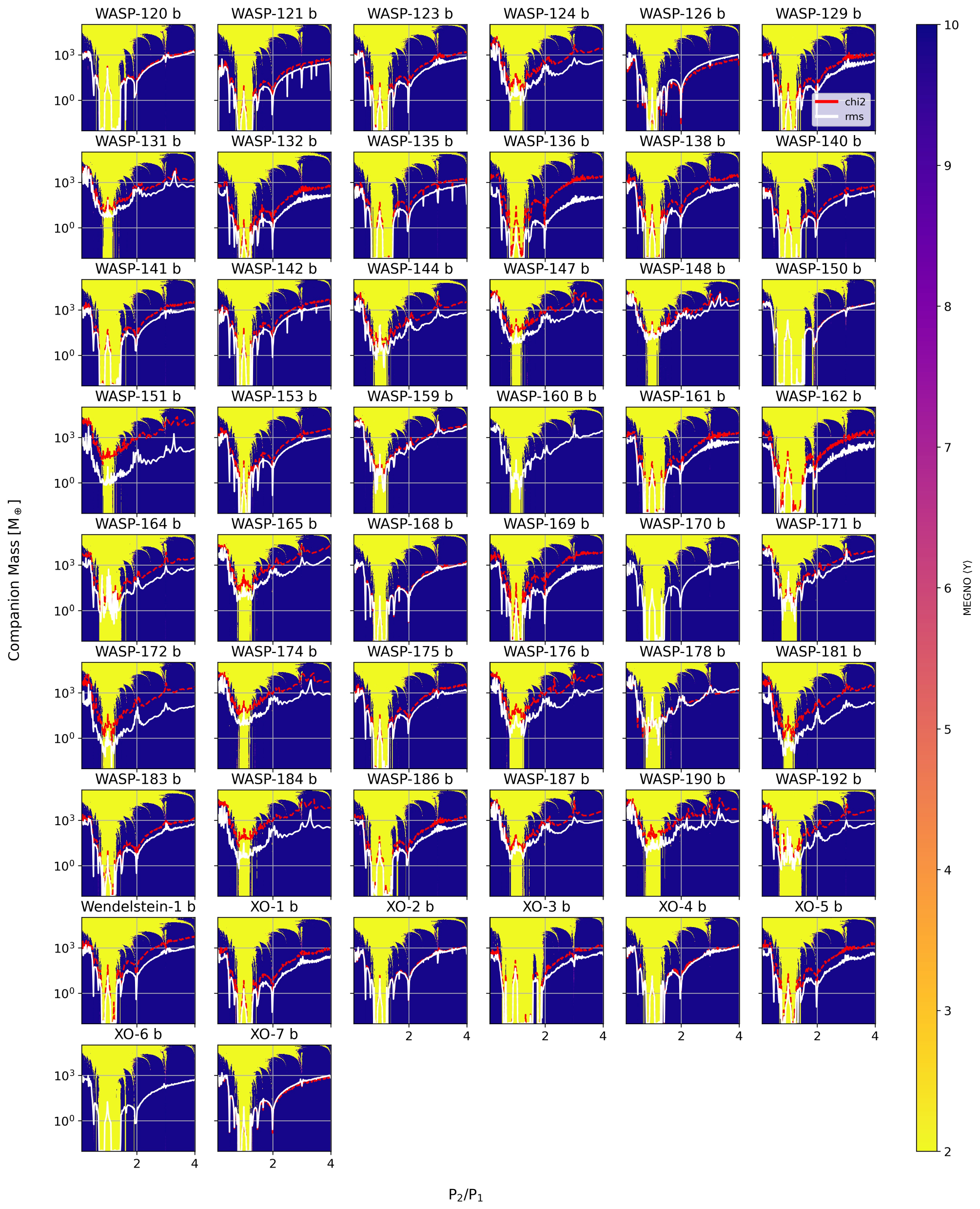}
    \caption{Upper mass limit of companion planets and MEGNO for 260 HJs in our sample. (Continued)}
    \figurenum{A1}
\end{figure*}

{
\section{Star and Hot Jupiter mass}
In this appendix, we provide the masses of the host stars and hot Jupiters for 260 systems included in our research.
\startlongtable
\begin{deluxetable}{ccccc}
	\tablecaption{The mass of the host star and the planet in the system. The mass of the star is in solar mass and the mass of the planet is in Jupiter mass. The references for the mass of the star and the planet are also provided. \label{tab:mass}}
	\tablehead{
		\colhead{Planet Name} & \colhead{$\mathrm{M_\star}\ \big[\mathrm{M_\odot}\big]$} & \colhead{$\mathrm{M_\star}$ Ref} & \colhead{$\mathrm{M_p}\ \big[\mathrm{M_J}\big]$} & \colhead{$\mathrm{M_p}$ Ref}
	}
	\startdata
		  CoRoT-1 b &               0.95 &                        \citet{Barge08} &             1.02979 & Calculated from MSINI and I \\
		  CoRoT-2 b &               0.97 &                       \citet{Alonso08} &             3.27527 & Calculated from MSINI and I \\
		  HAT-P-1 b &              1.133 &                       \citet{Torres08} &            0.532369 & Calculated from MSINI and I \\
		  HAT-P-2 b &              1.308 &                       \citet{Torres08} &             8.87066 & Calculated from MSINI and I \\
		  HAT-P-3 b &              0.928 &                       \citet{Torres08} &            0.595712 & Calculated from MSINI and I \\
		  HAT-P-4 b &              1.248 &                       \citet{Torres08} &            0.671507 & Calculated from MSINI and I \\
		  HAT-P-5 b &              1.157 &                       \citet{Torres08} &             1.05561 & Calculated from MSINI and I \\
		  HAT-P-6 b &               1.29 &                       \citet{Torres08} &             1.05957 & Calculated from MSINI and I \\
		  HAT-P-7 b &                1.5 &        \citet{Christensen-Dalsgaard10} &              1.7916 & Calculated from MSINI and I \\
		  HAT-P-8 b &               1.28 &                       \citet{Latham-09} &             1.29336 & Calculated from MSINI and I \\
		  HAT-P-9 b &               1.28 &                      \citet{Shporer09} &            0.776601 & Calculated from MSINI and I \\
		 HAT-P-13 b &               1.22 &                        \citet{Bakos09} &             0.85708 & Calculated from MSINI and I \\
		 HAT-P-14 b &              1.386 &                       \citet{Torres-10} &             2.23603 & Calculated from MSINI and I \\
		 HAT-P-16 b &              1.218 &                     \citet{Buchhave10} &              4.2025 & Calculated from MSINI and I \\
		 HAT-P-19 b &              0.842 &                      \citet{Hartman11} &            0.292273 & Calculated from MSINI and I \\
		 HAT-P-20 b &              0.756 &                        \citet{2011ApJ...742..116B} &             7.28953 & Calculated from MSINI and I \\
		 HAT-P-21 b &              0.947 &                        \citet{2011ApJ...742..116B} &             4.07833 & Calculated from MSINI and I \\
		 HAT-P-22 b &              0.916 &                        \citet{2011ApJ...742..116B} &             2.15137 & Calculated from MSINI and I \\
		 HAT-P-23 b &               1.13 &                        \citet{2011ApJ...742..116B} &             2.09606 & Calculated from MSINI and I \\
		 HAT-P-24 b &              1.191 &                      \citet{2010ApJ...725.2017K} &            0.685857 & Calculated from MSINI and I \\
		 HAT-P-25 b &               1.01 &                        \citet{2012ApJ...745...80Q} &            0.567181 & Calculated from MSINI and I \\
		 HAT-P-27 b &               0.92 &                        \citet{2012ApJ...760..139B} &            0.617485 & Calculated from MSINI and I \\
		 HAT-P-28 b &              1.025 &                     \citet{2011ApJ...733..116B} &            0.627833 & Calculated from MSINI and I \\
		 HAT-P-29 b &              1.207 &                     \citet{2011ApJ...733..116B} &            0.779004 & Calculated from MSINI and I \\
		 HAT-P-30 b &              1.242 &                      \citet{2011ApJ...735...24J} &            0.711088 & Calculated from MSINI and I \\
		 HAT-P-31 b &              1.218 &                      \citet{2011AJ....142...95K} &             2.16904 & Calculated from MSINI and I \\
		 HAT-P-32 b &               1.16 &                      \citet{2011ApJ...742...59H} &            0.861516 & Calculated from MSINI and I \\
		 HAT-P-33 b &              1.375 &                      \citet{2011ApJ...742...59H} &            0.763256 & Calculated from MSINI and I \\
		 HAT-P-34 b &              1.392 &                        \citet{2012AJ....144...19B} &               3.334 & Calculated from MSINI and I \\
		 HAT-P-35 b &              1.236 &                        \citet{2012AJ....144...19B} &             1.05422 & Calculated from MSINI and I \\
		 HAT-P-36 b &              1.022 &                        \citet{2012AJ....144...19B} &             1.83923 & Calculated from MSINI and I \\
		 HAT-P-37 b &              0.929 &                        \citet{2012AJ....144...19B} &             1.17362 & Calculated from MSINI and I \\
		 HAT-P-38 b &              0.886 &                         \citet{2012PASJ...64...97S} &            0.267569 & Calculated from MSINI and I \\
		 HAT-P-39 b &              1.404 &                      \citet{2012AJ....144..139H} &            0.599196 & Calculated from MSINI and I \\
		 HAT-P-40 b &              1.512 &                      \citet{2012AJ....144..139H} &              0.6202 & Calculated from MSINI and I \\
		 HAT-P-41 b &              1.418 &                      \citet{2012AJ....144..139H} &            0.800254 & Calculated from MSINI and I \\
		 HAT-P-42 b &               1.18 &                \citet{2013AA...558A..86B} &               1.044 & \citet{2013AA...558A..86B} \\
		 HAT-P-43 b &               1.03 & \citet{2022MNRAS.509.1075C} &               0.662 & \citet{2013AA...558A..86B} \\
		 HAT-P-44 b &              0.942 &                      \citet{2014AJ....147..128H} &               0.352 &           \citet{2014AJ....147..128H} \\
		 HAT-P-49 b &              1.543 &                      \citet{2014AJ....147...84B} &             1.72959 & Calculated from MSINI and I \\
		 HAT-P-50 b &               1.27 &               \citet{2015AJ....150..168H} &                1.35 & \citet{2015AJ....150..168H} \\
		 HAT-P-51 b &               0.98 &               \citet{2015AJ....150..168H} &               0.309 & \citet{2015AJ....150..168H} \\
		 HAT-P-52 b &               0.89 &               \citet{2015AJ....150..168H} &               0.818 & \citet{2015AJ....150..168H} \\
		 HAT-P-55 b &               1.01 &               \citet{2015PASP..127..851J} &               0.582 & \citet{2015PASP..127..851J} \\
		 HAT-P-56 b &              1.296 &                  \citet{2015AJ....150...85H} &             2.18029 & Calculated from MSINI and I \\
		 HAT-P-58 b &               1.03 &                 \citet{2021AJ....162....7B} &               0.372 & \citet{2021AJ....162....7B} \\
		 HAT-P-59 b &               1.01 &                 \citet{2021AJ....162....7B} &                1.54 & \citet{2021AJ....162....7B} \\
		 HAT-P-60 b &               1.44 &                 \citet{2021AJ....162....7B} &               0.574 & \citet{2021AJ....162....7B} \\
		 HAT-P-61 b &                1.0 &                 \citet{2021AJ....162....7B} &               1.057 & \citet{2021AJ....162....7B} \\
		 HAT-P-62 b &               1.02 &                 \citet{2021AJ....162....7B} &               0.761 & \citet{2021AJ....162....7B} \\
		 HAT-P-65 b &               1.21 &               \citet{2016AJ....152..182H} &               0.527 & \citet{2016AJ....152..182H} \\
		 HAT-P-66 b &               1.26 &               \citet{2016AJ....152..182H} &               0.783 & \citet{2016AJ....152..182H} \\
		 HAT-P-67 b &               1.64 &                  \citet{2017AJ....153..211Z} &                0.34 & \citet{2017AJ....153..211Z} \\
		 HAT-P-68 b &               0.68 &                \citet{2021AJ....161...64L} &               0.724 & \citet{2021AJ....161...64L} \\
		 HAT-P-69 b &               1.65 &                  \citet{2019AJ....158..141Z} &                3.58 & \citet{2019AJ....158..141Z} \\
		 HAT-P-70 b &               1.89 &                  \citet{2019AJ....158..141Z} &                6.78 & \citet{2019AJ....158..141Z} \\
		   HATS-1 b &              0.986 &                \citet{2013AJ....145....5P} &             1.86507 & Calculated from MSINI and I \\
		   HATS-2 b &              0.882 &               \citet{2013AA...558A..55M} &             1.34915 & Calculated from MSINI and I \\
		   HATS-3 b &              1.209 &                \citet{2013AJ....146..113B} &              1.0735 & Calculated from MSINI and I \\
		   HATS-4 b &              1.001 &                \citet{2014AJ....148...29J} &               1.323 &                 \citet{2014AJ....148...29J} \\
		   HATS-6 b &              0.574 &                      \citet{2015AJ....149..166H} &               0.319 &                \citet{2015AJ....149..166H} \\
		  HATS-13 b &              0.962 &      \citet{2015AA...580A..63M} &            0.542398 & Calculated from MSINI and I \\
		  HATS-16 b &               0.97 &      \citet{2016PASP..128g4401C} &                3.27 &         \citet{2016PASP..128g4401C} \\
		  HATS-18 b &               1.04 &      \citet{2019AJ....158...39C} &              1.9795 &   \citet{2019AJ....158...39C} \\
		  HATS-22 b &               0.76 &      \citet{2017MNRAS.468..835B} &                2.74 &           \citet{2017MNRAS.468..835B}\\
		  HATS-23 b &               1.12 &      \citet{2017MNRAS.468..835B} &                1.47 &           \citet{2017MNRAS.468..835B} \\
		  HATS-24 b &               1.07 &      \citet{2019RNAAS...3...35O} &                2.26 &        \citet{2019RNAAS...3...35O} \\
		  HATS-25 b &               0.99 &      \citet{2016AJ....152..108E} &               0.613 &        \citet{2016AJ....152..108E} \\
		  HATS-26 b &                1.3 &      \citet{2016AJ....152..108E} &                0.65 &        \citet{2016AJ....152..108E} \\
		  HATS-27 b &               1.42 &      \citet{2016AJ....152..108E} &                0.53 &        \citet{2016AJ....152..108E} \\
		  HATS-28 b &               0.93 &      \citet{2016AJ....152..108E} &               0.672 &        \citet{2016AJ....152..108E} \\
		  HATS-30 b &               1.09 &      \citet{2016AJ....152..108E} &               0.706 &        \citet{2016AJ....152..108E}\\
		  HATS-31 b &               1.28 &      \citet{2016AJ....152..161D} &                0.88 &    \citet{2016AJ....152..161D} \\
		  HATS-32 b &                1.1 &      \citet{2016AJ....152..161D} &                0.92 &    \citet{2016AJ....152..161D}\\
		  HATS-33 b &               1.06 &      \citet{2016AJ....152..161D} &               1.192 &    \citet{2016AJ....152..161D}\\
		  HATS-34 b &               0.96 &      \citet{2016AJ....152..161D} &               0.941 &    \citet{2016AJ....152..161D}\\
		  HATS-39 b &               1.38 &      \citet{2018MNRAS.477.3406B} &                0.63 &           \citet{2018MNRAS.477.3406B}\\
		  HATS-40 b &               1.56 &      \citet{2018MNRAS.477.3406B} &                1.59 &           \citet{2018MNRAS.477.3406B}\\
		  HATS-41 b &                1.5 &      \citet{2018MNRAS.477.3406B} &                 9.7 &           \citet{2018MNRAS.477.3406B}\\
		  HATS-42 b &               1.27 &      \citet{2018MNRAS.477.3406B} &                1.88 &           \citet{2018MNRAS.477.3406B}\\
		  HATS-43 b &               0.84 &      \citet{2018AJ....155..112B} &               0.261 &           \citet{2018AJ....155..112B}\\
		  HATS-44 b &               0.86 &      \citet{2018AJ....155..112B} &                0.56 &           \citet{2018AJ....155..112B}\\
		  HATS-45 b &               1.27 &      \citet{2018AJ....155..112B} &                 0.7 &           \citet{2018AJ....155..112B}\\
		  HATS-47 b &               0.67 &      \citet{2020AJ....159..173H} &               0.369 &         \citet{2020AJ....159..173H}\\
		  HATS-51 b &               1.19 &      \citet{2018AJ....155...79H} &               0.768 &         \citet{2018AJ....155...79H}\\
		  HATS-52 b &               1.11 &      \citet{2018AJ....155...79H} &                2.24 &         \citet{2018AJ....155...79H}\\
		  HATS-53 b &               0.96 &      \citet{2018AJ....155...79H} &               0.595 &         \citet{2018AJ....155...79H}\\
		  HATS-54 b &               1.05 &      \citet{2023MNRAS.518.4845J} &               0.753 &         \citet{2023MNRAS.518.4845J}\\
		  HATS-55 b &                1.2 &      \citet{2019AJ....158...63E} &               0.921 &        \citet{2019AJ....158...63E}\\
		  HATS-56 b &               1.57 &      \citet{2019AJ....158...63E} &               0.602 &        \citet{2019AJ....158...63E}\\
		  HATS-57 b &               1.03 &      \citet{2019AJ....158...63E} &               3.147 &        \citet{2019AJ....158...63E}\\
		  HATS-59 b &               1.04 &      \citet{2018AJ....156..216S} &               0.806 &          \citet{2018AJ....156..216S}\\
		  HATS-60 b &                1.1 &      \citet{2019AJ....157...55H} &               0.662 &         \citet{2019AJ....157...55H}\\
		  HATS-63 b &               0.93 &      \citet{2019AJ....157...55H} &                0.96 &         \citet{2019AJ....157...55H}\\
		  HATS-64 b &               1.56 &      \citet{2019AJ....157...55H} &                0.96 &         \citet{2019AJ....157...55H}\\
		  HATS-67 b &               1.44 &      \citet{2019AJ....157...55H} &                1.45 &         \citet{2019AJ....157...55H}\\
		  HATS-68 b &               1.35 &      \citet{2019AJ....157...55H} &                1.29 &         \citet{2019AJ....157...55H}\\
		  HATS-69 b &               0.89 &      \citet{2019AJ....157...55H} &               0.577 &         \citet{2019AJ....157...55H}\\
		  HATS-70 b &               1.78 &      \citet{2019AJ....157...31Z} &                12.9 &            \citet{2019AJ....157...31Z}\\
		  HATS-71 b &               0.49 &      \citet{2020AJ....159..267B} &                0.37 &           \citet{2020AJ....159..267B}\\
		HD 189733 b &              0.806 &    \citet{2008ApJ...677.1324T} &              1.1436 & Calculated from MSINI and I \\
		   K2-237 b &               1.26 &     \citet{2020AJ....160..209I} &               1.363 &      \citet{2020AJ....160..209I} \\
		   KELT-3 b &              1.282 &     \citet{2013ApJ...773...64P} &             1.46164 & Calculated from MSINI and I \\
		   KELT-6 b &              1.085 &     \citet{2014AJ....147...39C} &             0.43122 & Calculated from MSINI and I \\
		   KELT-7 b &              1.535 &     \citet{2015AJ....150...12B} &                1.28 &                \citet{2015AJ....150...12B} \\
		   KELT-8 b &               0.81 &     \citet{2017AJ....153..136S} &                0.66 &         \citet{2017AJ....153..136S} \\
		   KELT-9 b &               2.52 &     \citet{2017Natur.546..514G} &                2.88 &           \citet{2017Natur.546..514G} \\
		  KELT-15 b &               2.06 &      \citet{2017AJ....153..136S} &                1.31 &         \citet{2017AJ....153..136S}\\
		  KELT-16 b &               1.21 &      \citet{2017AJ....153...97O} &                2.75 &          \citet{2017AJ....153...97O}\\
		  KELT-17 b &               1.64 &      \citet{2016AJ....152..136Z} &                1.31 &            \citet{2016AJ....152..136Z}\\
		  KELT-18 b &               1.52 &      \citet{2017AJ....153..263M} &                1.18 &          \citet{2017AJ....153..263M}\\
		  KELT-20 b &               1.76 &      \citet{2017AJ....154..194L} &               3.382 &            \citet{2017AJ....154..194L}\\
		  KELT-21 b &               1.46 &      \citet{2018AJ....155..100J} &                3.91 &         \citet{2018AJ....155..100J}\\
		  KELT-24 b &               1.46 &      \citet{2019AJ....158..197R} &                5.18 &       \citet{2019AJ....158..197R}\\
			KPS-1 b &               0.89 &    \citet{2018PASP..130g4401B} &                1.09 &        \citet{2018PASP..130g4401B} \\
		MASCARA-4 b &               1.75 &    \citet{2020AA...635A..60D} &                 3.1 &          \citet{2020AA...635A..60D} \\
		   NGTS-1 b &               0.62 &     \citet{2018MNRAS.475.4467B} &               0.812 &         \citet{2018MNRAS.475.4467B} \\
		   NGTS-2 b &               1.64 &     \citet{2018MNRAS.481.4960R} &                0.74 &         \citet{2018MNRAS.481.4960R} \\
		 NGTS-3 A b &               1.02 &       \citet{2018MNRAS.478.4720G} &                2.38 &     \citet{2018MNRAS.478.4720G} \\
		   NGTS-6 b &               0.77 &    \citet{2019MNRAS.489.4125V} &               1.339 &           \citet{2019MNRAS.489.4125V} \\
		   NGTS-9 b &               1.34 &     \citet{2020MNRAS.491.2834C} &                 2.9 &          \citet{2020MNRAS.491.2834C} \\
	  OGLE-TR-113 b &              0.779 &      \citet{2008ApJ...677.1324T} &                1.26 &                 \citet{2008ApJ...677.1324T} \\
		  Qatar-1 b &               0.85 &      \citet{2011MNRAS.417..709A} &             1.08979 & Calculated from MSINI and I \\
		  Qatar-6 b &               0.82 &      \citet{2018AJ....155...52A} &               0.668 &         \citet{2018AJ....155...52A}\\
		  Qatar-7 b &               1.41 &      \citet{2019AJ....157...74A} &                1.88 &         \citet{2019AJ....157...74A}\\
		  Qatar-8 b &               1.03 &      \citet{2019AJ....157..224A} &               0.371 &         \citet{2019AJ....157..224A}\\
		  Qatar-9 b &               0.72 &      \citet{2019AJ....157..224A} &                1.19 &         \citet{2019AJ....157..224A}\\
		 Qatar-10 b &               1.16 &       \citet{2019AJ....157..224A} &               0.736 &         \citet{2019AJ....157..224A} \\
		   TrES-1 b &              0.878 &     \citet{2008ApJ...677.1324T} &            0.752484 & Calculated from MSINI and I \\
		   TrES-2 b &              0.983 &     \citet{2008ApJ...677.1324T} &             1.20071 & Calculated from MSINI and I \\
		   TrES-3 b &              0.915 &     \citet{2008ApJ...677.1324T} &                1.91 &               \citet{2009ApJ...691.1145S} \\
		   TrES-4 b &              1.394 &     \citet{2008ApJ...677.1324T} &               0.925 &               \citet{2009ApJ...691.1145S} \\
		   TrES-5 b &              0.893 &     \citet{2011ApJ...741..114M} &              1.7781 & Calculated from MSINI and I \\
		   WASP-1 b &                1.2 &     \citet{2010AA...516A..33E} &               0.918 &                \citet{2011MNRAS.414.3023S} \\
		   WASP-2 b &               0.88 &     \citet{2010AA...516A..33E} &            0.908659 & Calculated from MSINI and I \\
		   WASP-3 b &               1.22 &     \citet{2010AA...516A..33E} &             2.01184 & Calculated from MSINI and I \\
		   WASP-4 b &               0.91 &     \citet{2010AA...516A..33E} &             1.22311 & Calculated from MSINI and I \\
		   WASP-5 b &               1.01 &     \citet{2010AA...516A..33E} &             1.62423 & Calculated from MSINI and I \\
		   WASP-6 b &               0.93 &     \citet{2010AA...516A..33E} &             0.52136 & Calculated from MSINI and I \\
		   WASP-7 b &                1.2 &     \citet{2010AA...516A..33E} &            0.919279 & Calculated from MSINI and I \\
		   WASP-8 b &               1.03 &     \citet{2010AA...517L...1Q} &             2.13797 &                         \citet{2010AA...517L...1Q} \\
		  WASP-10 b &               0.79 &      \citet{2010AA...516A..33E} &             3.19144 & Calculated from MSINI and I \\
		  WASP-11 b &                0.8 &      \citet{2010AA...516A..33E} &            0.539755 & Calculated from MSINI and I \\
		  WASP-12 b &               1.28 &      \citet{2010AA...516A..33E} &             1.36063 & Calculated from MSINI and I \\
		  WASP-13 b &               1.09 &      \citet{2012MNRAS.419.1248B} &            0.473883 & Calculated from MSINI and I \\
		  WASP-14 b &               1.31 &      \citet{2010AA...516A..33E} &             7.69234 & Calculated from MSINI and I \\
		  WASP-15 b &               1.18 &      \citet{2010AA...516A..33E} &            0.542797 & Calculated from MSINI and I \\
		  WASP-16 b &                1.0 &      \citet{2010AA...516A..33E} &            0.842294 & Calculated from MSINI and I \\
		  WASP-17 b &               1.19 &      \citet{2010AA...516A..33E} &            0.508547 & Calculated from MSINI and I \\
		  WASP-18 b &               1.22 &      \citet{2010AA...516A..33E} &             10.2006 & Calculated from MSINI and I \\
		  WASP-19 b &               0.93 &      \citet{2010AA...516A..33E} &             1.13339 & Calculated from MSINI and I \\
		  WASP-20 b &                1.2 &      \citet{2015AA...575A..61A} &               0.311 &               \citet{2015AA...575A..61A}\\
		  WASP-21 b &               1.01 &      \citet{2010AA...519A..98B} &            0.300272 & Calculated from MSINI and I \\
		  WASP-22 b &                1.1 &      \citet{2010AJ....140.2007M} &            0.559059 & Calculated from MSINI and I \\
		  WASP-23 b &               0.78 &      \citet{2011yCat..35310024T} &            0.872226 & Calculated from MSINI and I \\
		  WASP-24 b &              1.184 &      \citet{2010ApJ...720..337S} &               1.091 & Calculated from MSINI and I \\
		  WASP-25 b &                1.0 &      \citet{2011MNRAS.410.1631E} &             0.57847 & Calculated from MSINI and I \\
		  WASP-26 b &               1.12 &      \citet{2010AA...520A..56S} &             1.01746 & Calculated from MSINI and I \\
		  WASP-31 b &              1.165 &      \citet{2011AA...531A..60A} &            0.479329 & Calculated from MSINI and I \\
		  WASP-32 b &               1.07 &      \citet{2012ApJ...760..139B} &             3.45362 & Calculated from MSINI and I \\
		  WASP-34 b &               1.01 &      \citet{2011AA...526A.130S} &            0.583354 & Calculated from MSINI and I \\
		  WASP-35 b &               1.07 &      \citet{2011AJ....142...86E} &            0.717098 & Calculated from MSINI and I \\
		  WASP-36 b &               1.02 &      \citet{2012AJ....143...81S} &             2.26929 & Calculated from MSINI and I \\
		  WASP-37 b &              0.925 &      \citet{2011AJ....141....8S} &             1.79362 & Calculated from MSINI and I \\
		  WASP-39 b &               0.93 &      \citet{2011AA...531A..40F} &            0.284338 & Calculated from MSINI and I \\
		  WASP-41 b &               0.93 &      \citet{2016AA...586A..93N} &            0.939101 & Calculated from MSINI and I \\
		  WASP-42 b &              0.881 &      \citet{2012AA...544A..72L} &            0.496968 & Calculated from MSINI and I \\
		  WASP-43 b &               0.58 &      \citet{2011AA...535L...7H} &             1.77609 & Calculated from MSINI and I \\
		  WASP-44 b &              0.951 &      \citet{2013yCat..74221988A} &            0.889836 & Calculated from MSINI and I \\
		  WASP-45 b &              0.909 &      \citet{2013yCat..74221988A} &             1.00657 & Calculated from MSINI and I \\
		  WASP-46 b &              0.956 &      \citet{2013yCat..74221988A} &             2.10232 & Calculated from MSINI and I \\
		  WASP-47 b &               1.02 &      \citet{2013AA...558A.106M} &             1.12341 & Calculated from MSINI and I \\
		  WASP-48 b &               1.19 &      \citet{2011AJ....142...86E} &            0.984236 & Calculated from MSINI and I \\
		  WASP-49 b &              0.938 &      \citet{2012AA...544A..72L} &            0.378231 & Calculated from MSINI and I \\
		  WASP-50 b &              0.892 &      \citet{2011yCat..35330088G} &             1.47244 & Calculated from MSINI and I \\
		  WASP-53 b &               0.84 &      \citet{2017MNRAS.467.1714T} &               2.132 &          \citet{2017MNRAS.467.1714T}\\
		  WASP-54 b &              1.213 &      \citet{2013AA...551A..73F} &             0.63273 & Calculated from MSINI and I \\
		  WASP-56 b &              1.107 &      \citet{2013AA...551A..73F} &            0.605385 & Calculated from MSINI and I \\
		  WASP-57 b &              0.954 &      \citet{2013AA...551A..73F} &            0.675887 & Calculated from MSINI and I \\
		  WASP-58 b &               0.94 &      \citet{2013AA...549A.134H} &            0.891379 & Calculated from MSINI and I \\
		  WASP-59 b &              0.719 &      \citet{2013AA...549A.134H} &            0.859095 & Calculated from MSINI and I \\
		  WASP-60 b &              1.078 &      \citet{2013AA...549A.134H} &            0.512143 & Calculated from MSINI and I \\
		  WASP-61 b &               1.22 &      \citet{2012MNRAS.426..739H} &             2.05491 & Calculated from MSINI and I \\
		  WASP-62 b &               1.25 &      \citet{2012MNRAS.426..739H} &             0.56227 & Calculated from MSINI and I \\
		  WASP-63 b &               1.32 &      \citet{2012MNRAS.426..739H} &            0.378096 & Calculated from MSINI and I \\
		  WASP-64 b &              1.004 &      \citet{2013AA...552A..82G} &             1.27137 & Calculated from MSINI and I \\
		  WASP-65 b &               0.93 &      \citet{2013AA...559A..36G} &                1.55 &           \citet{2013AA...559A..36G}\\
		  WASP-66 b &                1.3 &      \citet{2012MNRAS.426..739H} &             2.31346 & Calculated from MSINI and I \\
		  WASP-69 b &              0.826 &      \citet{2014MNRAS.445.1114A} &            0.259543 & Calculated from MSINI and I \\
		  WASP-71 b &              1.572 &      \citet{2013AA...552A.120S} &               2.242 &                  \citet{2013AA...552A.120S}\\
		  WASP-72 b &              1.327 &      \citet{2013AA...552A..82G} &             1.40771 & Calculated from MSINI and I \\
		  WASP-73 b &               1.34 &      \citet{2014yCat..35630143D} &             1.88059 & Calculated from MSINI and I \\
		  WASP-78 b &               1.33 &      \citet{2012AA...547A..61S} &            0.883997 & Calculated from MSINI and I \\
		  WASP-79 b &               1.52 &      \citet{2012AA...547A..61S} &            0.887514 & Calculated from MSINI and I \\
		  WASP-80 b &               0.58 &      \citet{2013yCat..35510080T} &            0.551763 & Calculated from MSINI and I \\
		  WASP-81 b &               1.08 &      \citet{2017MNRAS.467.1714T}&               0.729 &          \citet{2017MNRAS.467.1714T}\\
		  WASP-82 b &               1.64 &      \citet{2015AcA....65..117S} &                1.25 &                  \citet{2015AcA....65..117S}\\
		  WASP-83 b &               1.11 &      \citet{2015AJ....150...18H} &                 0.3 &                \citet{2015AJ....150...18H}\\
		  WASP-84 b &              0.842 &      \citet{2014MNRAS.445.1114A} &             0.69431 & Calculated from MSINI and I \\
		  WASP-87 b &                1.2 &      \citet{2016ApJ...823...29A} &                2.18 &         \citet{2016ApJ...823...29A}\\
		  WASP-90 b &               1.55 &      \citet{2016AA...585A.126W} &                0.63 &            \citet{2016AA...585A.126W}\\
		  WASP-91 b &               0.84 &      \citet{2017AA...604A.110A} &                1.34 &        \citet{2017AA...604A.110A}\\
		  WASP-92 b &               1.19 &      \citet{2016MNRAS.463.3276H} &               0.805 &             \citet{2016MNRAS.463.3276H}\\
		  WASP-93 b &               1.33 &      \citet{2016MNRAS.463.3276H} &                1.47 &             \citet{2016MNRAS.463.3276H}\\
		 WASP-104 b &              1.076 &       \citet{2014AA...570A..64S} &               1.272 &                  \citet{2014AA...570A..64S} \\
		 WASP-105 b &               0.89 &       \citet{2017AA...604A.110A} &                 1.8 &        \citet{2017AA...604A.110A} \\
		 WASP-106 b &              1.092 &       \citet{2014AA...570A..64S} &               1.925 &                  \citet{2014AA...570A..64S} \\
		 WASP-113 b &               1.32 &       \citet{2016AA...593A.113B} &               0.475 &          \citet{2016AA...593A.113B} \\
		 WASP-118 b &               1.32 &       \citet{2016MNRAS.463.3276H} &               0.514 &             \citet{2016MNRAS.463.3276H} \\
		 WASP-119 b &               1.02 &       \citet{2016AA...591A..55M} &                1.23 &          \citet{2016AA...591A..55M} \\
		 WASP-120 b &               1.39 &       \citet{2016PASP..128f4401T} &                4.85 &          \citet{2016PASP..128f4401T} \\
		 WASP-121 b &               1.36 &       \citet{2020AA...635A.205B} &               1.157 &        \citet{2020AA...635A.205B} \\
		 WASP-123 b &               1.17 &       \citet{2016PASP..128f4401T} &               0.899 &          \citet{2016PASP..128f4401T} \\
		 WASP-124 b &               1.07 &       \citet{2016AA...591A..55M} &                 0.6 &          \citet{2016AA...591A..55M} \\
		 WASP-126 b &               1.12 &       \citet{2019AJ....158..243P} &             0.28411 &                \citet{2019AJ....158..243P} \\
		 WASP-129 b &                1.0 &       \citet{2016AA...591A..55M} &                 1.0 &          \citet{2016AA...591A..55M} \\
		 WASP-131 b &               1.06 &       \citet{2017MNRAS.465.3693H} &                0.27 &         \citet{2017MNRAS.465.3693H} \\
		 WASP-132 b &               0.78 &       \citet{2022AJ....164...13H} &                0.41 &         \citet{2022AJ....164...13H} \\
		 WASP-135 b &               0.98 &       \citet{2016PASP..128b4401S} &             1.90043 & Calculated from MSINI and I \\
		 WASP-136 b &               1.41 &       \citet{2017AA...599A...3L} &                1.51 &             \citet{2017AA...599A...3L} \\
		 WASP-138 b &               1.22 &       \citet{2017AA...599A...3L} &                1.22 &             \citet{2017AA...599A...3L} \\
		 WASP-140 b &                0.9 &       \citet{2017MNRAS.465.3693H} &                2.44 &         \citet{2017MNRAS.465.3693H} \\
		 WASP-141 b &               1.25 &       \citet{2017MNRAS.465.3693H} &                2.69 &         \citet{2017MNRAS.465.3693H} \\
		 WASP-142 b &               1.33 &       \citet{2017MNRAS.465.3693H} &                0.84 &         \citet{2017MNRAS.465.3693H} \\
		 WASP-144 b &               0.81 &       \citet{2019MNRAS.482.1379H} &                0.44 &         \citet{2019MNRAS.482.1379H} \\
		 WASP-147 b &               1.04 &       \citet{2019MNRAS.482..301L} &               0.275 &           \citet{2019MNRAS.482..301L} \\
		 WASP-148 b &               0.95 &       \citet{2022AA...663A.134A} &               0.287 &        \citet{2022AA...663A.134A} \\
		 WASP-150 b &               1.39 &       \citet{2020yCat..51590255C} &                8.46 &           \citet{2020yCat..51590255C} \\
		 WASP-151 b &               1.08 &       \citet{2020PASP..132a4401M} &               0.316 &     \citet{2020PASP..132a4401M} \\
		 WASP-153 b &               1.34 &       \citet{2018AA...610A..63D} &                0.39 &       \citet{2018AA...610A..63D} \\
		 WASP-159 b &               1.41 &       \citet{2019MNRAS.482.1379H} &                0.55 &         \citet{2019MNRAS.482.1379H} \\
	   WASP-160 B b &               0.87 &     \citet{2019MNRAS.482..301L} &               0.278 &           \citet{2019MNRAS.482..301L} \\
		 WASP-161 b &               1.39 &       \citet{2019AJ....157...43B} &                2.49 &        \citet{2019AJ....157...43B} \\
		 WASP-162 b &               0.95 &       \citet{2019MNRAS.482.1379H} &                 5.2 &         \citet{2019MNRAS.482.1379H} \\
		 WASP-164 b &               0.95 &       \citet{2019MNRAS.482..301L} &                2.13 &           \citet{2019MNRAS.482..301L} \\
		 WASP-165 b &               1.25 &       \citet{2019MNRAS.482..301L} &               0.658 &           \citet{2019MNRAS.482..301L} \\
		 WASP-168 b &               1.08 &       \citet{2019MNRAS.482.1379H} &                0.42 &         \citet{2019MNRAS.482.1379H} \\
		 WASP-169 b &               1.34 &       \citet{2019MNRAS.489.2478N} &               0.561 &         \citet{2019MNRAS.489.2478N} \\
		 WASP-170 b &               0.93 &       \citet{2019AJ....157...43B} &                 1.6 &        \citet{2019AJ....157...43B} \\
		 WASP-171 b &               1.17 &       \citet{2019MNRAS.489.2478N} &               1.084 &         \citet{2019MNRAS.489.2478N} \\
		 WASP-172 b &               1.49 &       \citet{2019MNRAS.482.1379H} &                0.47 &         \citet{2019MNRAS.482.1379H} \\
		 WASP-174 b &               1.24 &       \citet{2020AA...633A..30M} &                0.33 &         \citet{2020AA...633A..30M} \\
		 WASP-175 b &               1.21 &       \citet{2019MNRAS.489.2478N} &                0.99 &         \citet{2019MNRAS.489.2478N} \\
		 WASP-176 b &               1.34 &       \citet{2020yCat..51590255C} &               0.855 &           \citet{2020yCat..51590255C} \\
		 WASP-178 b &               2.07 &       \citet{2019MNRAS.490.1479H} &                1.66 &         \citet{2019MNRAS.490.1479H} \\
		 WASP-181 b &               1.04 &       \citet{2019MNRAS.485.5790T} &               0.299 &          \citet{2019MNRAS.485.5790T} \\
		 WASP-183 b &               0.78 &       \citet{2019MNRAS.485.5790T} &               0.502 &          \citet{2019MNRAS.485.5790T} \\
		 WASP-184 b &               1.23 &       \citet{2019MNRAS.490.1479H} &                0.57 &         \citet{2019MNRAS.490.1479H} \\
		 WASP-186 b &               1.22 &       \citet{2020MNRAS.499..428S} &                4.22 &        \citet{2020MNRAS.499..428S} \\
		 WASP-187 b &               1.54 &       \citet{2020MNRAS.499..428S} &                 0.8 &        \citet{2020MNRAS.499..428S} \\
		 WASP-190 b &               1.35 &       \citet{2019AJ....157..141T} &                 1.0 &          \citet{2019AJ....157..141T} \\
		 WASP-192 b &               1.09 &       \citet{2019MNRAS.490.1479H} &                 2.3 &         \citet{2019MNRAS.490.1479H} \\
	Wendelstein-1 b &               0.65 &    \citet{2020AA...639A.130O} &               0.592 &       \citet{2020AA...639A.130O} \\
			 XO-1 b &              1.027 &       \citet{2008ApJ...677.1324T} &            0.918464 & Calculated from MSINI and I \\
			 XO-2 b &              0.974 &       \citet{2008ApJ...677.1324T} &            0.566541 & Calculated from MSINI and I \\
			 XO-3 b &               1.41 &       \citet{2008ApJ...677..657J} &             13.2849 & Calculated from MSINI and I \\
			 XO-4 b &                1.1 &       \citet{2017AJ....153..136S} &                1.42 &         \citet{2017AJ....153..136S} \\
			 XO-5 b &                1.0 &       \citet{2008ApJ...686.1331B} &             1.15251 & Calculated from MSINI and I \\
			 XO-6 b &               1.47 &       \citet{2017AJ....153...94C} &                 4.4 &         \citet{2017AJ....153...94C} \\
			 XO-7 b &               1.41 &       \citet{2020AJ....159...44C} &               0.709 &         \citet{2020AJ....159...44C} \\
	\enddata

	\end{deluxetable}

}


\label{lastpage}
\end{document}